\documentstyle[12pt,twoside]{article}
\begin{document}

\newif\iffigs\figsfalse
\figstrue

\iffigs
  \input epsf
\else
  \message{No figures will be included. See TeX file for more
information.}
\fi

\thispagestyle{empty}
\begin{flushright}
MPI-PhT/98-78
\end{flushright}
\bigskip\bigskip\bigskip\begin{center}
{\LARGE {Correction to four-loop RG functions}}
\vskip 7pt
{\LARGE {in the two-dimensional lattice O($n$) $\sigma$-model}}
\end{center}  
\vskip 1.0truecm
\centerline{Dong-Shin Shin}
\vskip5mm
\centerline{Max-Planck-Institut f\"ur
 Physik}
\centerline{ -- Werner-Heisenberg-Institut -- }
\centerline{F\"ohringer Ring 6, 80805 Munich, Germany}
\vskip 2cm
\bigskip \nopagebreak \begin{abstract}
\noindent
We report the result of our evaluation of the Feynman diagrams appearing
in the determination of the four-loop renormalization group
functions in the two-dimensional lattice O($n$) $\sigma$-model 
by Caracciolo and Pelissetto.
In the list of the integrals used for the determination of those functions,
we find that three entries were not correctly evaluated.
We give the values for them corrected by us including those
for all other integrals which we computed with higher precision.
These results are then 
applied to revise the determination of the second analytic
correction to correlation length $\xi$ and spin susceptibility $\chi$
by Caracciolo et al. as well as our determination of the mass gap 
by means of a finite volume technique
where we explicitly made use of the four-loop $\beta$-function.
In both cases we find sizeable changes in predictions.
In the meantime there appeared a paper by Alles et al. where they revised
one finite integral in the list of our corrected integrals.
After having taken the new revised value into consideration, we found that
there are no noticeable changes in the perturbative predictions 
of the present paper including the final conclusions.
\end{abstract}
\vskip 1.5cm

\newpage\setcounter{page}1

\def\eqb{\begin{equation}}
\def\eqe{\end{equation}}
\def\dib{\begin{displaymath}}
\def\die{\end{displaymath}}
\def\eqnb{\begin{eqnarray}}
\def\eqne{\end{eqnarray}}
\def\eqsb{\begin{eqnarray*}}
\def\eqse{\end{eqnarray*}}
\def\md{{\rm d}}
\def\me{{\rm e}}

\section{Introduction}

The two-dimensional non-linear O($n$) $\sigma$-model 
is interesting in theoretical physics in many respects.
In condensed matter physics,
the model can be applied to the study of ferromagnetic systems. 
In elementary particle physics, on the other hand,
it serves as a good toy model for the investigation of the strong interactions
since it shares with QCD many common properties.

The most interesting property of the $\sigma$-model from the view of particle
physicists is that of asymptotic freedom 
which is predicted by perturbation theory.
Besides this, the model shows many additional interesting properties.
One believes a non-perturbative generation of a mass gap
via dimensional transmutation
which controls the exponential decay of correlation functions
at large distances.
From the existence of an infinite set of conserved non-local 
quantum charges~\cite{Lu1,Lu2,Bu1,Bu2},
one can show that there is no particle production in the theory
and also derive ``factorization equations'' directly.
These properties lead finally to the exact determination of the $S$-matrix
up to a CDD-ambiguity~\cite{Za1,Za2},
assuming that the spectrum of the theory is known.
By making use of the exact $S$-matrix and the thermodynamic Bethe ansatz,
Hasenfratz and Niedermayer determined 
the ratio of the mass gap to the $\Lambda$-parameter $m/\Lambda$
of the theory exactly~\cite{HaMaNi,HaNi}.
If one formulates the $\sigma$-model on the lattice, it has a further
advantage: due to low dimensionality it is simple to simulate numerically
and can therefore be used in many cases 
for testing new ideas in lattice theories.

The property of asymptotic freedom 
is based on the validity of perturbation theory.
One usually formulates perturbation theory on a non-symmetric trivial vacuum.
The Mermin-Wagner theorem~\cite{MeWa}, however, 
tells us that in O($n$) $\sigma$-model 
the continuous O($n$) symmetry can not be spontaneously broken
in two dimensions.
Therefore, the validity of the perturbation theory in infinite volume
is in fact not guaranteed.
This observation, together with the lack of clear asymptotic scaling 
in simulations,
gives some authors~\cite{PaSe1,PaSe2,PaSe3} a motive for an argument that
the non-linear O($n$) $\sigma$-model
has no mass gap and for any $n\geq 2$ it undergoes a
Kosterlitz-Thouless phase transition~\cite{KoTh,Ko}
at a finite coupling $f_c\not=0$.

Monte Carlo simulations allow in principle to explore the behavior of a model
in all areas of the bare coupling.
It is of general interest to connect Monte Carlo results 
with those of the perturbative 
renormalization group (RG) which should dictate the scaling laws of 
the theory near the fixed point $f_0=0$.
For a precise comparison with the Monte Carlo data, 
the computation of the RG functions at
high orders in perturbation theory is desirable. 
Until now they are known up to four loops where
the long and difficult four-loop computation was recently done by
Caracciolo and Pelissetto~\cite{CaPe}.
These determinations of the four-loop $\beta$- and $\gamma$-functions then
allow us to compute 
the correlation length $\xi$ and spin susceptibility $\chi$
up to second analytic correction.

We applied an efficient coordinate space method for the evaluation
of lattice Feynman diagrams described in detail in our paper~\cite{Sh1}
to compute the integrals listed in the paper by Caracciolo et al.
more precisely.
In most integrals, our results showed good agreement 
with those by Caracciolo et al. 
We, however, found that three integrals were not correctly evaluated.
Since these errors turn out to 
cause sizable changes in the perturbative prediction to
$\xi$ and $\chi$ as well as in our determination of 
the mass gap by a finite volume technique~\cite{Sh2},
we would like to report our results for the evaluation of the lattice integrals
together with the corrections to the physical prediction
to $\xi$ and $\chi$ as well as to our determination of the mass gap.
Furthermore, the four-loop RG functions evaluated by Caracciolo et al.
were applied in many papers~\cite{ACD,AlBu,BuCo,MePe,PaSe3}
and it is necessary to revise them in accordance with
our corrected values for the RG functions.

We would like to stress that we checked 
only a minor part of the four-loop RG functions by Caracciolo et al.,
where we just reevaluated the integrals listed in that paper.
We still do not know whether the other analytic formulas are correct.
For the complete confirmation of the four-loop RG functions, 
it would take a while until we finish our own computation of them
which is in progress~\cite{Sh3}.

The paper is organized as follows.
In the next section we define the $\sigma$-model 
and give in this model the hitherto known results for the perturbative 
RG functions including the four-loop determination
by Caracciolo et al. and its corrected version by us.
Section~\ref{sec3} is then devoted to the reinvestigation of the physical 
quantities by making use of our corrected values for the RG functions.
Finally, the explicit evaluations of lattice Feynman diagrams are
presented in the Appendix.

\section{Computation of the RG functions}\label{sec2}

We consider the non-linear $\mbox{O}(n)$ $\sigma$-model in two dimensions
defined by the action
\begin{equation}
S = \frac{1}{2f_0}\int \md^2x\Big[\partial_{\mu}q(x)\cdot\partial_{\mu}q(x)
-2 hq_n(x)\Big],
\end{equation}
where $q_i(x)$ with $i=1,\cdots,n$ is an $n$-component real scalar field
of unit length $q(x)^2=1$ and
$f_0$ denotes the bare coupling constant.
$h$ is here a magnetic field which regularizes the infrared divergences
and explicitly breaks the O($n$) symmetry, lining up the field along the
component $q_n$.
At small coupling, one can then consider small fluctuations around
the direction of the magnetic field. In this way,
the perturbative expansions in $f_0$ can be obtained by setting
\eqb
q(x)=\bigg\{\pi_1(x),\cdots,\pi_{n-1}(x),
\bigg[1-\sum_{\alpha=1}^{n-1}\pi_{\alpha}(x)^2\bigg]^{1/2}\bigg\}
\eqe
which solves the constraint $q(x)^2=1$.
The regulator $h$ can be removed at the final stage of the computation.
This is guarenteed by 
the conjecture of Elitzur~\cite{El}, proved later by David
order by order in perturbation theory~\cite{Da},
which states that in O($n$) $\sigma$-model,
O($n$) invariant correlation functions have a finite infrared limit
in two dimensions.

With this prescription, 
one can now start to compute the physically interesting quantities
according to perturbation theory.
Higher loop computations are however accompanied by 
ultraviolet divergences, which requires a renormalization.
This can be achieved by introducing a regularization and redefining
bare fields $\pi$ and coupling $f_0$ as renormalized ones, $\pi_R$ and $f$. 
In the dimensional regularization with dimension $d$,
one sets $\pi_R=Z^{-1/2}\pi$ and $f=Z_1^{-1}f_0\mu^{d-2}$, where the
$Z$ and $Z_1$, taken as a series in $f$, are adjusted so that 
the renormalized theory satisfies a chosen set of two independent conditions.
Choosing $\mu$ as the scale of the renormalized theory,
if one changes $\mu$, one has to change the renormalized fields and 
coupling accordingly, which is represented by the RG equation
\eqb\label{rgeqms}
\bigg\{\mu\frac{\partial}{\partial\mu}+
\beta(f)\frac{\partial}{\partial f}+\frac{1}{2}n\gamma(f)
+\bigg[\frac{1}{2}\gamma(f)+\frac1f\beta(f)\bigg]
h\frac{\partial}{\partial h}\bigg\}
G_R^{(n)}(p,f,\mu ;h) = 0,
\eqe
where $G_R^{(n)}$
is the renormalized $n$-point correlation function for the $\pi_R$ field
and $\beta(f)$ and $\gamma(f)$ are $\beta$- and $\gamma$-functions 
respectively:
\eqnb
\beta(f)&=&\mu\frac{\md}{\md\mu}f=
-b_1f^2-b_2f^3-b_3f^4-b_4f^5+{\cal O}(f^6),\\
\gamma(f)&=&\mu\frac{\md}{\md\mu}\ln Z=
c_1f+c_2f^2+c_3f^3+c_4f^4+{\cal O}(f^5).
\eqne
The coefficients $b_1$, $b_2$ and $c_1$ are universal in the 
sense that they do not depend on the renormalization scheme chosen 
and explicitly given by
\eqnb\label{beta1ms}
b_1 &=& \frac{n-2}{2\pi}, \\\label{beta2ms}
b_2 &=& \frac{n-2}{(2\pi)^2}, \\
c_1 &=& \frac{n-1}{2\pi}.
\eqne
All other higher terms are, on the other hand, scheme dependent.
In the minimal subtraction (MS) scheme, 
they were computed up to four loops~\cite{BrHi,Hi,BeWe,We}:
\eqnb
b_3 &=& \frac14\frac{n^2-4}{(2\pi)^3},\\
b_4 &=& \frac{n-2}{(2\pi)^4}
\bigg[-\frac{1}{12}(n^2-22n+34)+\frac32\zeta(3)(n-3)\bigg],\\
c_2 &=& 0, \\
c_3 &=& \frac{3}{32\pi^3}(n-1)(n-2),\\
c_4 &=& \frac{1}{192\pi^4}(n-1)(n-2)[4(5-n)+3(3-n)\zeta(3)].
\eqne
with $\zeta(3)\simeq 1.2020569$. 

If one formulates the $\sigma$-model on the lattice,
there are many different ways to define actions to
regularize the theory.
Here, we would like to consider the standard nearest-neighbor action 
\begin{equation}\label{actionth}
S = -\frac{1}{f_0}\sum_{x,\mu}q(x)\cdot q(x+\hat\mu) - h\sum_x q_n(x) .
\end{equation}
The RG equation on the lattice which is formally 
the same as Eq.~(\ref{rgeqms}),
but written in terms of the bare $\beta$- and $\gamma$-functions,
is given by
\eqb\label{rgeqlatt}
\bigg\{\hat\beta(f_0)\frac{\partial}{\partial f_0}
-a\frac{\partial}{\partial a}-\frac12n\hat\gamma(f_0)
+\bigg[\frac{1}{2}\hat\gamma(f_0)+\frac{1}{f_0}\hat\beta(f_0)\bigg]
h\frac{\partial}{\partial h}\bigg\}G_L^{(n)}(p,f_0,1/a;h) = 0,
\eqe
where $G_L^{(n)}$ is the lattice $n$-point correlation function
for the $\pi$ field and the lattice 
$\beta$- and $\gamma$-functions are given by
\eqnb
\hat\beta(f_0) &=& -a\frac{\md}{\md a}f_0=
-\hat b_1f_0^2-\hat b_2f_0^3-\hat b_3f_0^4-
\hat b_4f_0^5+{\cal O}(f_0^6),\\
\hat\gamma(f_0) &=& a\frac{\md}{\md a}\ln Z=
\hat c_1f_0+\hat c_2f_0^2+\hat c_3f_0^3+\hat c_4f_0^4+{\cal O}(f_0^5).
\eqne
The coefficients in this regularization are also known to four loops.
Up to three loops, they are given by~\cite{FaTr,Wz,CaPe2}
\eqnb\label{beta1}
\hat b_1 &=& b_1,\\\label{beta2}
\hat b_2 &=& b_2,\\\label{beta3}
\hat b_3 &=& \frac{n-2}{(2\pi)^3}
\bigg[\bigg(\frac12+\frac18\pi^2-4\pi^2G_1\bigg)(n-2)+1+\frac12\pi-\frac{5}{24}
\pi^2\bigg],\\
\hat c_1 &=& c_1,\\
\hat c_2 &=& \frac{n-1}{8\pi},\\
\hat c_3 &=& \frac{n-1}{(2\pi)^3}
\bigg[\bigg(\frac12-\frac18\pi^2+4\pi^2G_1\bigg)(n-2)+\frac{11}{24}\pi^2\bigg]
\eqne
with $G_1\simeq 0.0461636$.
Recently, Caracciolo and Pelissetto~\cite{CaPe} computed the quite demanding
four-loop coefficients $\hat b_4$ and $\hat c_4$:
\eqnb
\hat b_4 &=& \frac{n-2}{2\pi}\bigg\{\frac{2n-7}{96}+
\frac{n^2-4n+5}{32\pi}+\frac{n}{16\pi^2}+\frac{5n-9}{16\pi^3} +
     \frac{73n-164}{96\pi^3}\zeta(3) \nonumber\\ 
& & -\frac{n-2}{2\pi}(n-2+3\pi)G_1-\frac{n-2}{\pi}(3-\pi)R +
     \frac{n-6}{24}J \nonumber\\ 
& & -\frac{n-2}{3}L_1 
+\frac18(n-3)(n-2)(K-2V_3) \nonumber\\ & & \label{bbeta4}
+\frac{n-2}{2}\Big[2V_1+(n-2)V_2+4V_4
-2V_{5}-2V_{6}-16W_{2}\Big]\bigg\}, \\
\hat c_4 &=& \frac{n-1}{2\pi}\bigg\{-\frac{5n-21}{192}+
\frac{(n-2)^2}{32\pi}+\frac{3(n-2)}{32\pi^2}+\frac{n-2}{16\pi^3}\nonumber\\& &
-\frac{n-2}{192\pi^3}(6n+37)\zeta(3)-\frac{n-2}{4\pi}\Big[2(n-2)-5\pi\Big]G_1
+\frac{n-2}{2\pi}(3-\pi)R \nonumber\\& &
-\frac{n-6}{48}J -\frac{n-2}{3}L_1 
-\frac{1}{16}(n-3)(n-2)(K-2V_3) \nonumber\\& & \label{beta4}
+\frac{n-2}{4}\Big[(n-3)V_1+2(n-2)V_2-4V_4
+2V_{5}+2V_{6}+16W_{2}\Big]\bigg\},
\eqne
where the integrals $G_1,R,J,K,L_1,V_1,\cdots,V_6,W_1$ and $W_2$ are defined
in Appendix~\ref{appa} and their numerical values evaluated by 
Caracciolo et al. are listed in the middle column of Table~\ref{4loopc}.
\renewcommand{\arraystretch}{1.2}
\begin{table}[htb]\centering
\begin{tabular}{c||c|c}\hline
Integrals & \hspace{0.4cm} Values I \hspace{0.7cm} & 
\hspace{0.4cm} Values II \hspace{2.7cm} \\\hline
$G_1$  & \phantom{-}0.0461636 
& \phantom{-}0.04616362923(1)\phantom{5741} \\
$R$    & \phantom{-}0.0148430 
& \phantom{-}0.014842965985741(1) \\
$J$    & \phantom{-}0.1366198 
& \phantom{-}0.136619772367581(1) \\
$K$    & \phantom{-}0.095887\phantom{8}  
& \phantom{-}0.09588764425(1)\phantom{7581} \\
$L_1$  & \phantom{-}0.0029334 
& \phantom{-}0.002933289310(1)\phantom{758} \\
$V_1$  & \phantom{-}0.016961\phantom{8} 
& \phantom{-}0.016961078576(1)\phantom{758} \\
$V_2$  &           -0.00114\phantom{81}             
&           -0.0011381683(1)\phantom{75814} \\
$V_3$  & \phantom{-}0.07243\phantom{81}  
& \phantom{-}0.07243631946(1)\phantom{7581} \\
$V_4$  &           -0.0013125           
&           -0.002624929(1)\phantom{758145} \\
$V_5$  & \phantom{-}0.010063\phantom{8}  
& \phantom{-}0.0100630371(1)\phantom{75814} \\
$V_6$  & \phantom{-}0.017507\phantom{8}  
& \phantom{-}0.017506946(1)\phantom{758145} \\
$W_1$  &           -0.0296860           
&           -0.007421482994(1)\phantom{751} \\
$W_2$  & \phantom{-}0.00221\phantom{81} 
& \phantom{-}0.0006923019(1)\phantom{75814} \\\hline
\end{tabular}
\parbox{13.5cm}
{\caption{\label{4loopc}
Numerical evaluations of the lattice integrals appearing in the 
\mbox{four-loop} $\beta$- and $\gamma$-functions. 
The numbers in the middle column \mbox{(Values I)} are 
the evaluations by Caracciolo et al., while the numbers 
in the right column (Values II)
are those by the author.}}
\end{table}

By applying the coordinate space method for the evaluation of lattice Feynman
diagrams described in detail in our paper~\cite{Sh1},
we reevaluated all integrals in Table~\ref{4loopc} with higher precision.
We show our results in the right column of the same table.
As we see in Table~\ref{4loopc},
we reproduce most values by Caracciolo et al.
with a minor precision mistake for the constant $L_1$.
However, for three integrals\footnote{Recently, 
Alles et al.~\cite{AlCaPe} revised the value
for $W_2$. We improved its numerical precision to get 
$W_2=0.0021225468(1)$~\cite{ShVe}.}
$V_4$, $W_1$ and $W_2$,
our numbers are different from those by Caracciolo et al.,
where the difference in $V_4$ can be accounted for by a factor of 2.

We remark that Caracciolo et al. checked their determination 
of the four-loop RG functions by the technique of $1/n$~expansion.
In the limit of large $n$, Eqs.~(\ref{bbeta4}) and (\ref{beta4}) reduce to
\eqnb\label{beta4ln}
\frac{\hat b_4}{n^3} &=& \frac{1}{2\pi^2}\bigg[\frac{1}{32}-
\frac{G_1}{2}+\frac{\pi}{8}(K+4V_2-2V_3)\bigg]+{\cal O}\bigg(\frac1n\bigg), 
\\\label{gamma4ln}
\frac{\hat c_4}{n^3} &=& \frac{1}{2\pi}\bigg[\frac{1}{32\pi}-
\frac{1}{32\pi^3}\zeta(3)-
\frac{G_1}{2\pi}-\frac{K}{16}+\frac{V_1}{4}+\frac{V_2}{2}+\frac{V_3}{8}\bigg]
+{\cal O}\bigg(\frac1n\bigg).
\eqne
Eqs.~(\ref{beta4ln}) and (\ref{gamma4ln}) agree with the predictions by the
$1/n$~expansion in the given order~\cite{CaPe}.
However, as we see in the above equations,
this method can not control all integrals in Table~\ref{4loopc},
merely those of $G_1,K,V_1,V_2$ and $V_3$.

There was, on the other hand,
a further check on the RG functions
when we determined a mass gap~\cite{Sh2} by applying a perturbative
computation of the mass gap $M(L)$ in finite volume $L$.
In converting our computation of $M(L)$ on the lattice to the renormalized
MS scheme, we made there an explicit use of the four-loop $\beta$-function.
From the constraint that we should formally have a free theory 
in the O(2) model with dimensional regularization,
we get a relationship for the perturbative $M(L)$ to satisfy.
In this case, we could however confirm only the integral $J$.

\section{Application to physical quantities}\label{sec3}
\setcounter{equation}{0}

In this section we present the application of the perturbative determination
of the RG functions in the previous section to the physical predictions,
explicitly the analytic correction to the correlation length $\xi$ 
and spin susceptibility $\chi$ 
as well as the determination of a mass gap through a perturbative
computation of a finite volume mass gap $M(L)$.
We first consider the application to the 
analytic correction to $\xi$ and $\chi$.

\subsection{Perturbative evaluation of $\xi$ and $\chi$}

The determination of the four-loop $\beta$- and $\gamma$-functions 
makes it possible to compute $\xi$ and $\chi$
up to second analytic correction.
We discuss here the numerical evaluation of the analytic corrections
to $\xi$ and $\chi$ and also to the other relevant quantities.
We first consider the correlation length $\xi$.

The correlation length is a physical quantity and correspondingly 
satisfies the RG equation:
\eqb\label{rgexi}
\bigg[\hat\beta(f_0)\frac{\partial}{\partial f_0}-
a\frac{\partial}{\partial a}\bigg]\xi(f_0,a) = 0
\eqe
We may then define the RG invariant parameter as
\eqnb\label{xig}
\xi/a &=& C_{\xi}(\hat b_1f_0)^{\hat b_2/\hat b_1^2}
\exp\bigg(\frac{1}{\hat b_1f_0}\bigg)\bar\xi(f_0), \\\label{bxig}
\bar\xi(f_0) &=& 
\exp\bigg[\int_0^{f_0}\md x\bigg(\frac{1}{\hat\beta(x)}+
\frac{1}{\hat b_1x^2}-\frac{\hat b_2}{\hat b_1^2x}\bigg)\bigg]
\eqne
which satisfies Eq.~(\ref{rgexi}) order by order in perturbation theory.
The constant $C_\xi$ is non-perturbative and its value depends on the explicit
definition of the correlation length.
For the isovector exponential correlation length $\xi_V^{({\rm exp})}$,
which controls the large distance behaviour of the two-point function
$\langle q(x)\cdot q(0)\rangle$, an exact determination was achieved 
by Hasenfratz and Niedermayer~\cite{HaMaNi,HaNi}
using the thermodynamic Bethe ansatz:
\eqb\label{xicn}
C_{\xi_V^{({\rm exp})}} = \bigg(\frac{\me}{8}\bigg)^{1/(n-2)}
\Gamma\bigg(1+\frac{1}{n-2}\bigg)2^{-5/2}\exp\bigg[-\frac{\pi}{2(n-2)}\bigg].
\eqe
An alternative definition with respect to the isovector correlation function
\mbox{$\langle q(x)\cdot q(0)\rangle$}
is the second-moment correlation length
\eqb
\xi_V^{(2)} = \sqrt{\frac{\sum_x|x|^2\langle q(x)\cdot q(0)\rangle}
{\sum_x\langle q(x)\cdot q(0)\rangle}}.
\eqe
There are also other definitions with respect to higher isospin channels,
e.g. the exponential correlation length related to the isotensor
correlation function \mbox{$\langle [q(x)\cdot q(0)]^2-1/n\rangle$}
and the corresponding second-moment correlation length.

The function $\bar\xi(f_0)$ of Eq.~(\ref{bxig})
is regular at $f_0=0$ and may be expanded 
in $f_0$:
\eqb
\bar\xi(f_0) = 1+A_1f_0+A_2f_0^2+{\cal O}(f_0^3),
\eqe
where
\eqnb\label{a1}
A_1 &=& \frac{\hat b_1\hat b_3-\hat b_2^2}{\hat b_1^3}, \\\label{a2}
A_2 &=& \frac{(\hat b_2^2-\hat b_1\hat b_3)^2
+\hat b_1^2(\hat b_2^3-2\hat b_1\hat b_2\hat b_3+\hat b_1^2\hat b_4)}
{2\hat b_1^6}.
\eqne
Inserting $\hat b_1,\hat b_2,\hat b_3$ and $\hat b_4$ from 
Eqs.~(\ref{beta1})-(\ref{beta3}) and (\ref{bbeta4})
in Eqs.~(\ref{a1}) and (\ref{a2}), we obtain for $A_1$ and $A_2$
\eqnb\label{lda1}
A_1 &=& 
\frac{1}{n-2}\bigg[\bigg(\frac{1}{4\pi}+\frac{\pi}{16}-2\pi G_1\bigg)(n-2)
+\frac14-\frac{5\pi}{48}\bigg], \\
A_2 &=& \frac{1}{192(n-2)^2}\bigg[\frac{\pi^2}{24}(3n-11)^2+\pi(4n^2-19n+17)
\nonumber\\& &
+6n^3-39n^2+95n-76\bigg]\nonumber\\& &
+\frac{1}{32\pi^2(n-2)}[7n-16+2\pi(n-1)]
+\frac{1}{96\pi^2}\frac{73n-164}{n-2}\zeta(3)\nonumber\\& &
+\frac{G_1}{24(n-2)}[-12(n-2)(n-3)-12\pi(3n-5)-(3n-11)\pi^2]\nonumber\\& &
+2\pi^2G_1^2-(3-\pi)R+\frac{\pi}{24}\frac{n-6}{n-2}J-\frac{\pi}{3}L_1+
\frac{\pi}{8}(n-3)(K-2V_3)\nonumber\\& &\label{lda2}
+\frac12\pi\Big[2V_1+(n-2)V_2+4V_4-2V_5-2V_6-16W_2\Big].
\eqne
Now, if we substitute for the constants in Eqs.~(\ref{lda1}) and (\ref{lda2})
the values in Table~\ref{4loopc} evaluated by us, 
we get for $A_2$ an estimation 
which is different from that by Caracciolo et al.:
\eqnb
A_1 &\simeq& \frac{1}{n-2}(-0.0490 - 0.0141n), \\
A_2 &\simeq& \frac{1}{(n-2)^2}(0.188 - 0.122n + 0.0405n^2 - 0.0129n^3).
\eqne
It is to be remarked that if $n$ increases, the difference with the 
determination by Caracciolo et al. becomes small,
which is of course
to be expected since the four-loop coefficient is correct in the
large $n$ limit.

Next, we would like to compute the analytic correction to
the isovector spin susceptibility
$\chi_V = \sum_x\langle q(x)\cdot q(0)\rangle$.
It satisfies the following RG equation:
\eqb\label{rgechi}
\bigg[\hat\beta(f_0)\frac{\partial}{\partial f_0}-
a\frac{\partial}{\partial a}+\hat\gamma(f_0)\bigg]\chi_V(f_0,a) = 0 .
\eqe
We may then define the RG invariant parameter as
\eqnb\label{chig}
\chi_V/a^2 &=& C_{\chi}(\hat b_1f_0)^{(2\hat b_2+\hat c_1\hat b_1)/\hat b_1^2}
\exp\bigg(\frac{2}{\hat b_1f_0}\bigg)\bar\chi(f_0), \\\label{chigb}
\bar\chi(f_0) &=&
\exp\bigg[\int_0^{f_0}\md x\bigg(\frac{2}{\hat\beta(x)}+
\frac{2}{\hat b_1x^2}-\frac{2\hat b_2}{\hat b_1^2x}
-\frac{\hat\gamma(x)}{\hat\beta(x)}
-\frac{\hat c_1}{\hat b_1x}\bigg)\bigg].
\eqne
Eq.~(\ref{chig}) with Eq.~(\ref{chigb}) 
satisfies the RG equation (\ref{rgechi}) 
order by order in perturbation theory.
Contrary to $C_\xi$ in Eq.~(\ref{xig}), there is no exact expression 
for the non-perturbative constant $C_\chi$ at present.
It is however known in the large $n$ limit~\cite{BiCaRo,CaRo}:
\eqb
C_{\chi} = \frac{\pi}{16}\bigg[1+\frac{1}{n}(4+3\gamma_C-\pi-3\gamma_E-7\ln 2)
+{\cal O}\bigg(\frac{1}{n^2}\bigg)\bigg],
\eqe
where $\gamma_E$ is Euler's constant and
$\gamma_C\simeq 0.4861007$.

The function $\bar\chi(f_0)$ of Eq.~(\ref{chigb})
is well-defined at $f_0=0$ and may be expanded 
in $f_0$:
\eqb
\bar\chi(f_0) = 1+B_1f_0+B_2f_0^2+{\cal O}(f_0^3)
\eqe
with
\eqnb\label{b1}
B_1 &=& \frac{2\hat b_1\hat b_3-2\hat b_2^2+\hat b_1^2\hat c_2-
\hat b_1\hat b_2\hat c_1}{\hat b_1^3},\\\label{b2}
B_2 &=& \frac{[2\hat b_2^2+\hat b_1\hat b_2\hat c_1
-\hat b_1(2\hat b_3+\hat b_1\hat c_2)]^2}{2\hat b_1^6}\nonumber\\& &
+\frac{2\hat b_2^3+\hat b_1\hat b_2^2\hat c_1
-\hat b_1\hat b_2(4\hat b_3+\hat b_1\hat c_2)
+\hat b_1^2(2\hat b_4-\hat b_3\hat c_1+\hat b_1\hat c_3)}{2\hat b_1^4}.
\eqne
Inserting $\hat b_1,\cdots,\hat b_4$ and $\hat c_1,\cdots,\hat c_3$
from Eqs.~(\ref{beta1})-(\ref{bbeta4})
in Eqs.~(\ref{b1}) and (\ref{b2}), we obtain for $B_1$ and $B_2$
\eqnb\label{ldb1}
B_1 &=& 
\frac{1}{2\pi(n-2)}\bigg[\frac12\pi(n+1)-1
+\frac14\pi^2\bigg(n-\frac{11}{3}\bigg)
-8\pi^2(n-2)G_1\bigg], \\
B_2 &=& \frac{1}{(n-2)^2}\bigg[\frac{\pi^2}{1152}(3n-11)^2
+\frac{\pi}{96}(7n^2-30n+17)\nonumber\\& &
+\frac{1}{96}(n-1)(3n^2-13n+51)\nonumber\\& &
-\frac{1}{8\pi}(2n-1)+\frac{1}{8\pi^2}(3n^2-13n+15)\bigg]
+\frac{1}{48\pi^2}\frac{73n-164}{n-2}\zeta(3)\nonumber\\& &
+\frac{G_1}{6(n-2)}\Big[6(3n-4)-6\pi(4n-5)-(3n-11)\pi^2\Big]
+8\pi^2G_1^2\nonumber\\& &
-2(3-\pi)R+\frac{\pi}{12}\frac{n-6}{n-2}J-\frac{2\pi}{3}L_1
+\frac{\pi}{4}(n-3)(K-2V_3)\nonumber\\& & \label{ldb2}
+\pi[2V_1+(n-2)V_2+4V_4-2V_5-2V_6-16W_2].
\eqne
Now, if we substitute for the constants in Eqs.~(\ref{ldb1}) and (\ref{ldb2})
our values in Table~\ref{4loopc}, we obtain
\eqnb
B_1 &\simeq& \frac{1}{n-2}(-0.1888 + 0.0626n), \\
B_2 &\simeq& \frac{1}{(n-2)^2}(0.4191 - 0.2688n + 0.0517n^2 - 0.0108n^3).
\eqne
Again, for $B_2$ we get a prediction 
which is different from that by Caracciolo et al.

Let us also consider the ratio $\chi_V/\xi^2$ which has an advantage of
being possible to compute an additional analytic correction. 
Writing
\eqb
\frac{\chi_V}{\xi^2} = 
R_V\bigg(\frac{2\pi}{n-2}\frac{1}{f_0}\bigg)^{-(n-1)/(n-2)}
\bigg(1+\sum_{i=1}^{\infty}C_if_0^i\bigg),
\eqe
where $R_V=C_{\chi}/C_{\xi}^2$ is a non-perturbative universal constant,
the expansion coefficients $C_1$, $C_2$ and $C_3$ have the expressions
\eqnb
C_1 &=& \frac{1}{4\pi}\frac{n-1}{n-2}(\pi-2), \\
C_2 &=& \frac{n-1}{(n-2)^2}
\bigg[(n-2)^2G_1-\frac{1}{96}(3n^2-23n+31)
+\frac{n-1}{8\pi^2}-\frac{2n-3}{8\pi}\bigg], \\
C_3 &=& \frac{n-1}{(n-2)^3}\bigg[-\frac{1}{384}(10n^3-79n^2+186n-137)
\nonumber\\& &
+\frac{1}{192\pi}(2n-3)(3n^2-26n+37)\nonumber\\& &
+\frac{1}{32\pi^2}(n-1)(3n-5)-\frac{1}{48\pi^3}(3n^3-18n^2+38n-27)\bigg]
\nonumber\\& &
\frac{n-1}{n-2}\bigg\{-\frac{1}{192\pi^3}(2n^2+57n-134)\zeta(3)\nonumber\\& &
-\frac{1}{48}(n-6)J+\frac{G_1}{4\pi}\Big[(5n-9)\pi-2(2n-3)\Big]\bigg\}
\nonumber\\& &
+(n-1)\bigg[\frac{1}{2\pi}(3-\pi)R-\frac{n-3}{16}(K-2V_3)\nonumber\\& &
+\frac{n-7}{12}V_1+\frac12(-2V_4+V_5+V_6+8W_2)\bigg].
\eqne
The numerical evaluations for these coefficients 
with our values in Table~\ref{4loopc} yield
\eqnb
C_1 &\simeq& \frac{n-1}{n-2}0.0908 , \\
C_2 &\simeq& \frac{n-1}{(n-2)^2}(-0.0316 - 0.0120n + 0.0149n^2) , \\
C_3 &\simeq& \frac{n-1}{(n-2)^3}
(0.0578 - 0.0564n + 0.0273n^2 - 0.0149n^3 + 0.0041n^4) .
\eqne
As expected,
for the expansion coefficient $C_3$ our determination does not agree
with the estimation by Caracciolo et al.

Finally, we would like to
consider the spin-$k$ susceptibility~\cite{CaPe} defined as
\eqb
\chi^{(k)} = \sum_x
\langle Y_k^{\alpha_1,\cdots,\alpha_k}(0)
        Y_k^{\alpha_1,\cdots,\alpha_k}(x)\rangle ,
\eqe
where
\eqb
Y_k^{\alpha_1,\cdots,\alpha_k} =
q_{\alpha_1}\cdots q_{\alpha_k} - {\rm ``Traces"}
\eqe
and the ``Traces'' must be such that $Y_k^{\alpha_1,\cdots,\alpha_k}$ is 
symmetric and traceless.
Then, from the standard RG arguments one has
\eqb
\chi^{(k)} = C_{\chi}^{(k)}\exp\bigg(\frac{4\pi}{n-2}\frac{1}{f_0}\bigg)
\bigg(\frac{2\pi}{n-2}\frac{1}{f_0}\bigg)^{-[2+k(n+k-2)]/(n-2)}
\bigg[1+\sum_{i=1}^{\infty}B_i^{(k)}f_0^i\bigg]
\eqe
with a non-universal constant $C_{\chi}^{(k)}$.
By including the ratio 
\eqb
\frac{\chi^{(k)}}{\xi^2} = 
R^{(k)}\bigg(\frac{2\pi}{n-2}\frac{1}{f_0}\bigg)^{-k(n+k-2)/(n-2)}
\bigg[1+\sum_{i=1}^{\infty}C_i^{(k)}f_0^i\bigg]
\eqe
and its expansion coefficients 
\eqnb
C_1^{(k)} &=& \frac{1}{4\pi}\frac{k(n+k-2)}{n-2}(\pi-2),\\
C_2^{(k)} &=& \frac{1}{2}\Big[c_1^{(k)}\Big]^2+
\frac{k(n+k-2)}{n-2}\bigg[(n-2)G_1-\frac{1}{8\pi}-\frac{1}{96}(3n-14)\bigg],
\eqne
one gets for the expansion coefficients
\eqnb
B_1^{(k)} &=& 2A_1+C_1^{(k)},\\
B_2^{(k)} &=& 2A_2+A_1^2+C_2^{(k)}+2A_1C_1^{(k)}
\eqne
which have the numerical values
\eqnb
B_1^{(k)} &\simeq& \frac{1}{n-2}\Big[-0.0980 - 0.0283n + 0.0908k(k+n-2)\Big],\\
B_2^{(k)} &\simeq& 
\frac{1}{(n-2)^2}\Big[0.37866 - 0.2428n + 0.08115n^2 - 0.0257n^3 \nonumber\\& &
+ k(k+n-2)(-0.0363 - 0.0187n + 0.0149n^2) \nonumber\\& &
+ 0.0041k^2(k+n-2)^2\Big].
\eqne
Here, our value of $B_2^{(k)}$ is not in agreement with 
that by Caracciolo et al.

\renewcommand{\arraystretch}{1.2}
\begin{table}[t]\centering
\begin{tabular}{cccccc}\hline
n & $\beta$ & $\xi\phantom{33333}$ & ${\cal R}^{(2)}$ & 
${\cal R}^{(3)}$ & ${\cal R}^{(4)}$\\\hline
\mbox{} 3 \mbox{} & \mbox{} 1.85 \mbox{} & 
\mbox{} 89.16(21)\phantom{333} \mbox{} & 
\mbox{} 0.743(2)\phantom{3} \mbox{} & 
\mbox{} 0.781(2)\phantom{3} \mbox{} & \mbox{} 0.822(2)\phantom{3} \mbox{} \\
3 & 2.25 & 1049(7)\phantom{3333.} & 
0.861(6)\phantom{3} & 0.897(6)\phantom{3} & 0.928(6)\phantom{3}\\
3 & 2.60 & 8569(92)\phantom{333.} & 0.901(10) & 0.934(10) & 0.958(10)\\
3 & 3.00 & 94.6(1.6)$\cdot 10^3$  & 0.930(16) & 0.959(16) & 0.977(17)\\
4 & 2.50 & 34.85(9)\phantom{333.} & 0.911(2)\phantom{3} & 
0.930(2)\phantom{3} & 0.949(2)\phantom{3}\\
4 & 2.80 & 86.07(37)\phantom{33.} & 0.927(4)\phantom{3} & 
0.945(4)\phantom{3} & 0.960(4)\phantom{3}\\
8 & 5.80 & 33.41(8)\phantom{333.} & 0.985(2)\phantom{3} & 
0.990(2)\phantom{3} & 0.994(2)\phantom{3}\\\hline
\end{tabular}
{\caption{\label{rat}
Ratio of Monte Carlo results to the theoretical perturbative predictions at
two, three and four loops for the correlation length $\xi$
in the O(3), O(4) and O(8) model evaluated for various values of $\beta$.}}
\end{table}

Now, it is interesting to compare the theory with ``experiment''.
We would like to take the correlation length and compute the ratio
\eqb
{\cal R}^{(l)}(\beta)=
\frac{\xi_{{\rm MC}}(\beta)}{\xi_{{\rm th}}^{(l)}(\beta)},
\eqe
where $\xi_{{\rm MC}}(\beta)$ is the Monte Carlo value for 
the correlation length
given as a function of the inverse bare coupling $\beta=1/f_0$
and $\xi_{{\rm th}}^{(l)}(\beta)$ is the theoretical $l$-loop 
prediction from Eqs.~(\ref{xig})-(\ref{xicn}).
In Table~\ref{rat} we show determinations of ${\cal R}^{(l)}$
for some selected values of $n$ and $\beta$
by inserting our new corrected coefficient of the four-loop $\beta$-function.
Compared to the old estimations by Caracciolo et al.,
the correction of the four-loop $\beta$-function 
causes to reduce the four-loop contribution to the perturbative
correlation length in all of the considered cases,
where the difference is maximum
in the O(3) model for $\beta=1.85$ and becomes small 
if $\beta$ or/and $n$ increase.
It follows that, in every case, the four-loop contributions to the correlation
length are smaller than Caracciolo et al. have estimated,
and so the discrepancies between the theoretical four-loop
prediction and ``experiment'' are larger than the estimation
by Caracciolo et al. 
Explicitly, at the largest $\beta$ values investigated,
they are 2.3\% for the O(3) model, 
4.0\% for the O(4) model and 0.6\% for the O(8) model
and there still remains some space for higher loop contributions. 

\subsection{Determination of a mass gap by finite volume method}

In our work~\cite{Sh2} on the determination of a mass gap $m$
by applying the perturbative computation of 
the mass gap $M(L)$ in finite volume $L$,
we made an explicit use of the four-loop $\beta$-function 
in converting our computation of $M(L)$ regularized on a lattice 
to the renormalized MS-scheme.
Since the four-loop coefficient has to be changed, our prediction for
the mass gap has to be corrected accordingly, and so we would like to 
present here the modifications caused by the change of the $\beta$-function.

To the determination of the physical
infinite volume mass gap $m$ from the finite volume mass gap $M(L)$,
we applied two different methods:
one method was the determination 
by matching the behavior of the finite volume 
mass gap $M(L)$ evaluated at small $L$ with that of the mass shift
$\delta_0\equiv [M(L)-m]/m$ known at large $L$
and the other was 
by means of a coupling
$\bar{g}^2(L)$ running with $L$.
We analyzed both methods again
by inserting our new corrected value of the $\beta$-function.

In the case of the determination of the mass gap by applying
the mass shift, 
we concluded that the method is not stable and so it is difficult
to make a reasonable estimation for the mass gap by this method. 
After we substituted our new corrected value 
of the four-loop $\beta$-function,
we see substantial changes,
but the whole picture including the final conclusion
does not improve the situation.
Therefore, we do not want to present here the evaluation of the mass gap 
by this method.
As for the determination of the mass gap by means of running coupling,
we however see interesting effects 
through the new corrected value and therefore
we would like to discuss in the following
the new evaluation of the mass gap by this method in detail.

\begin{figure}[hp]
\begin{center}
\leavevmode
\epsfxsize=100mm
\epsfbox{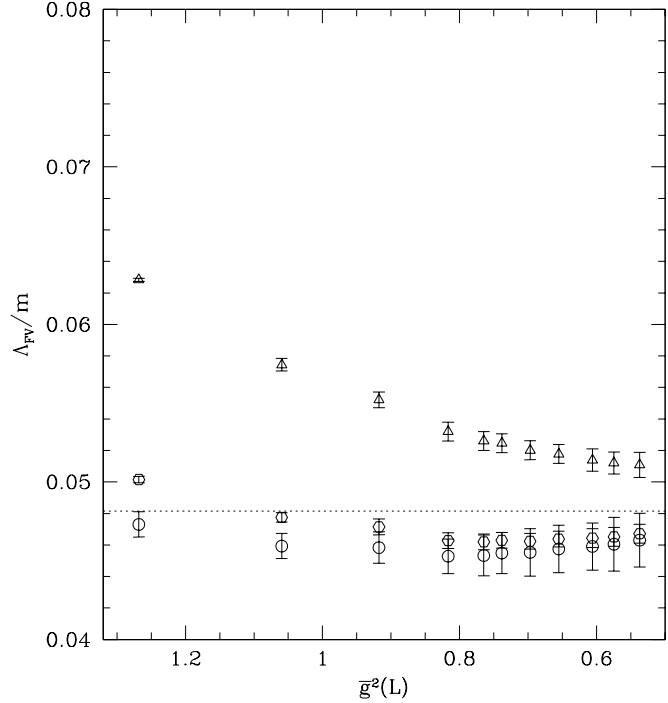}
\end{center}
\caption{
Two-, three- and four-loop approximations to $\Lambda_{FV}/m$ in the O(3) model
from Table~\ref{lambapp}.
The triangular points show the 2-loop approximation
to the $\Lambda$-parameter, the hexagonal points the 3-loop
and the circular points the 4-loop approximation.
The dotted line is the value of Hasenfratz et al. 
($\Lambda_{FV}/m=0.04815\cdots$).
}
\label{fvml}
\end{figure}

In Ref.~\cite{Sh2}, 
we introduced a coupling $\bar{g}^2(L)$ running with volume $L$ 
defined by\footnote{The running coupling $\bar{g}^2(L)$ was first defined 
in Ref.~\cite{LWW}.}
\begin{equation}\label{defrunningcoup}
\bar{g}^2(L) = 2M(L)L/(n-1),
\end{equation}
where $M(L)$ is a mass gap in finite volume 
which we computed to three-loop order in perturbation theory.
For the coefficients of $\beta$-function defined by
\begin{equation}
\tilde{\beta}(\bar{g}^2) \equiv -L\frac{\partial\bar{g}^2}{\partial L} 
  = -\bar{g}^2\sum_{l=1}^{\infty}\tilde{b}_l(\bar{g}^2)^l ,
\end{equation}
we found
\begin{eqnarray}\label{tilb1}
\tilde{b}_1 &=& b_1, \\\label{tilb2}
\tilde{b}_2 &=& b_2, \\\label{tilb3}
\tilde{b}_3 &=& \frac{(n-1)(n-2)}{(2\pi)^3}, \\
\tilde{b}_4 &=& \frac{1}{4}\frac{n-2}{(2\pi)^4}
\Big[(n-2)^3\chi_1 + (n-2)^2\chi_2 + (n-2)\chi_3 + \chi_4\Big]. \label{tilb4}
\end{eqnarray}
The explicit expressions for the constants $\chi_1,\cdots,\chi_4$
can be found in appendix~B of Ref.~\cite{Sh2},
where the values for $t_1$, $t_2$ and $t_3$ in their expressions
have now to be modified by our new values:
\eqb
\hat{b}_4 = \frac{n-2}{(2\pi)^4}\Big[(n-2)^2t_1+(n-2)t_2+t_3\Big]
\eqe
with\footnote{In the evaluation of $t_1$, $t_2$ and $t_3$ we took the revised
value for $W_2$ in Ref.~\cite{AlCaPe} into consideration.}
\begin{eqnarray}
t_1 &=& -1.015069687473002(1),\\
t_2 &=& -5.9168085(1),\\
t_3 &=& -9.0937547(1).
\end{eqnarray}
By inserting these corrected numbers into the expressions for 
$\chi_1,\cdots,\chi_4$, we get the numerical values
\begin{eqnarray}
\chi_1 &=& -1.20(1),\hspace{0.5cm}
\chi_2 = -3.63(1),\\
\chi_3 &=& \phantom{-}23.6(1),\\
\chi_4 &=& -5.2123414(1).
\end{eqnarray}
The $\Lambda$-parameter in this scheme is given by
\begin{eqnarray} \label{lambdal}
\Lambda_{FV} &=& \frac{1}{L}(b_1\bar{g}^2)^{-b_2/b_1^2}e^{-1/(b_1\bar{g}^2)}
  \cdot\tilde{\lambda}(\bar{g}^2) , \\\label{lambllamb}
\tilde{\lambda}(\bar{g}^2) &=& \exp\bigg[-\int_0^{\bar{g}^2} dx
  \bigg(\frac{1}{\tilde{\beta}(x)}+\frac{1}{b_1x^2}
-\frac{b_2}{b_1^2x}\bigg)\bigg]
\end{eqnarray}
which has the following perturbative expansion 
up to order ${\cal O}(\bar{g}^4)$:
\begin{eqnarray}\label{2looplambdafv}
\Lambda^{(2)}_{FV} &=& 
  \frac{1}{L}(b_1\bar{g}^2)^{-b_2/b_1^2}e^{-1/(b_1\bar{g}^2)}, \\
\Lambda^{(3)}_{FV} &=& 
  \frac{1}{L}(b_1\bar{g}^2)^{-b_2/b_1^2}e^{-1/(b_1\bar{g}^2)}
  \bigg\{1+\frac{b_2^2-b_1\tilde{b}_3}{b_1^3}\bar{g}^2\bigg\}, \\
\label{4looplambdafv}
\Lambda^{(4)}_{FV} &=& 
  \frac{1}{L}(b_1\bar{g}^2)^{-b_2/b_1^2}e^{-1/(b_1\bar{g}^2)}
  \bigg\{1+\frac{b_2^2-b_1\tilde{b}_3}{b_1^3}\bar{g}^2 \nonumber\\ & &
  +\frac{b_2^4-b_1^2b_2^3+2b_1^3b_2\tilde{b}_3-2b_1b_2^2\tilde{b}_3
  +b_1^2\tilde{b}_3^2-b_1^4\tilde{b}_4}{2b_1^6}\bar{g}^4\bigg\}.
\end{eqnarray}
If we insert the running couplings $\bar{g}^2(L)$ 
from Table~1 of Ref.~\cite{Sh2},
measured for given $L$ 
in units of the mass gap $m$ in O(3) model,
in Eqs.~(\ref{2looplambdafv})-(\ref{4looplambdafv}),
we obtain new values for the ratio $\Lambda_{FV}/m$ in 2-, 3- and 
4-loop approximations which are listed in Table~\ref{lambapp}
and also illustrated in Fig.~\ref{fvml}.

The new effect which mainly arises
by substituting our corrected value for the
four-loop $\beta$-function is that,
compared to the old diagram with the value by Caracciolo et al.,
for large couplings the new four-loop estimations
for $\Lambda_{FV}/m$ become smaller, so that
the four-loop curve in Fig.~\ref{fvml} shifts underneath.
As a result, the order of 3- and 4-loop curves changes.
However, when the coupling goes to zero, the difference with the old diagram
becomes smaller and smaller, and at the smallest coupling considered,
it is practically zero.
From this observation we can conclude that the changes in $\beta$-function
influence only at large couplings and for small couplings,
which are indeed the region of our main interest, 
we have the same picture as the earlier investigation with the value
by Caracciolo et al.
Therefore, as in the case with the old four-loop 
$\beta$-function, our determination of the 
mass gap converges to the value predicted 
by Hasenfratz and Niedermayer very well.

To give a further impression on the systematic errors, we also consider
an alternative definition of $n$-loop approximations to $\Lambda_{FV}$;
instead of expanding $\tilde{\lambda}(\bar{g}^2)$ of Eq.~(\ref{lambllamb})
in the coupling $\bar{g}^2$,
we insert the perturbative coefficients of 
$\tilde{\beta}(x)$ in $\tilde{\lambda}(\bar{g}^2)$ and integrate this exactly.
If we apply the coefficients of the $\beta$-function
calculated up to 4-loop order, 
we obtain the following approximations to $\Lambda_{FV}$ at
2-, 3- and 4-loop levels
which are a little modified from
Eqs.(\ref{2looplambdafv})-(\ref{4looplambdafv}):
\begin{eqnarray} \label{lambdam}
\bar{\Lambda}^{(2)}_{FV} &=& 
 \frac{1}{L}(b_1\bar{g}^2)^{-b_2/b_1^2}e^{-1/(b_1\bar{g}^2)} \nonumber\\ & &
 \cdot\exp\bigg[-\int_0^{\bar{g}^2}dx
 \bigg(\frac{1}{x^2(b_1+b_2x)}+\frac{1}{b_1x^2}-\frac{b_2}{b_1^2x}\bigg)\bigg],
\\
\bar{\Lambda}^{(3)}_{FV} &=& 
  \frac{1}{L}(b_1\bar{g}^2)^{-b_2/b_1^2}e^{-1/(b_1\bar{g}^2)} \nonumber\\ & &
  \cdot\exp\bigg[-\int_0^{\bar{g}^2}dx
  \bigg(\frac{1}{x^2(b_1+b_2x+\tilde{b}_3x^2)}
  +\frac{1}{b_1x^2}-\frac{b_2}{b_1^2x}\bigg)\bigg], \\
\bar{\Lambda}^{(4)}_{FV} &=& 
  \frac{1}{L}(b_1\bar{g}^2)^{-b_2/b_1^2}e^{-1/(b_1\bar{g}^2)} 
\nonumber\\\label{lambdam4} & &
  \cdot\exp\bigg[-\int_0^{\bar{g}^2}dx
  \bigg(\frac{1}{x^2(b_1+b_2x+\tilde{b}_3x^2+\tilde{b}_4x^3)}
  +\frac{1}{b_1x^2}-\frac{b_2}{b_1^2x}\bigg)\bigg].
\end{eqnarray}

\begin{table}[tp]\centering
\begin{tabular}{cccc}\hline
\hspace{0.5cm} \mbox{} $\bar{g}^2(L)$ \mbox{} \hspace{0.5cm} & 
\hspace{0.5cm} \mbox{} $\Lambda^{(2)}_{FV}/m$ \mbox{} \hspace{0.5cm} &  
\hspace{0.5cm} \mbox{} $\Lambda^{(3)}_{FV}/m$ \mbox{} \hspace{0.5cm} &  
\hspace{0.5cm} \mbox{} $\Lambda^{(4)}_{FV}/m$ \mbox{} \hspace{0.5cm} 
\\ \hline
 0.5372  &  0.0511(8)  &   0.0467(6)  &   0.0463(17)  \\
 0.5747  &  0.0512(7)  &   0.0465(6)  &   0.0460(17)  \\
 0.6060  &  0.0514(7)  &   0.0464(6)  &   0.0459(15)  \\
 0.6553  &  0.0518(6)  &   0.0464(5)  &   0.0457(15)  \\
 0.6970  &  0.0520(6)  &   0.0462(5)  &   0.0455(15)  \\
 0.7383  &  0.0525(6)  &   0.0463(5)  &   0.0455(13)  \\
 0.7646  &  0.0526(6)  &   0.0462(5)  &   0.0453(13)  \\
 0.8166  &  0.0532(6)  &   0.0463(5)  &   0.0453(11)  \\
 0.9176  &  0.0552(5)  &   0.0471(5)  &   0.0458(10)  \\
 1.0595  &  0.0574(4)  &   0.0478(3)  &   0.0459(8)\phantom{0}  \\
 1.2680  &  0.0628(1)  &   0.0502(2)  &   0.0473(8)\phantom{0}
\\ \hline
\end{tabular}
\caption{\label{lambapp}
Two-, three- and four-loop approximations to $\Lambda_{FV}/m$
in the O(3) model}
\end{table}

\begin{figure}[p]
\begin{center}
\leavevmode
\epsfxsize=100mm
\epsfbox{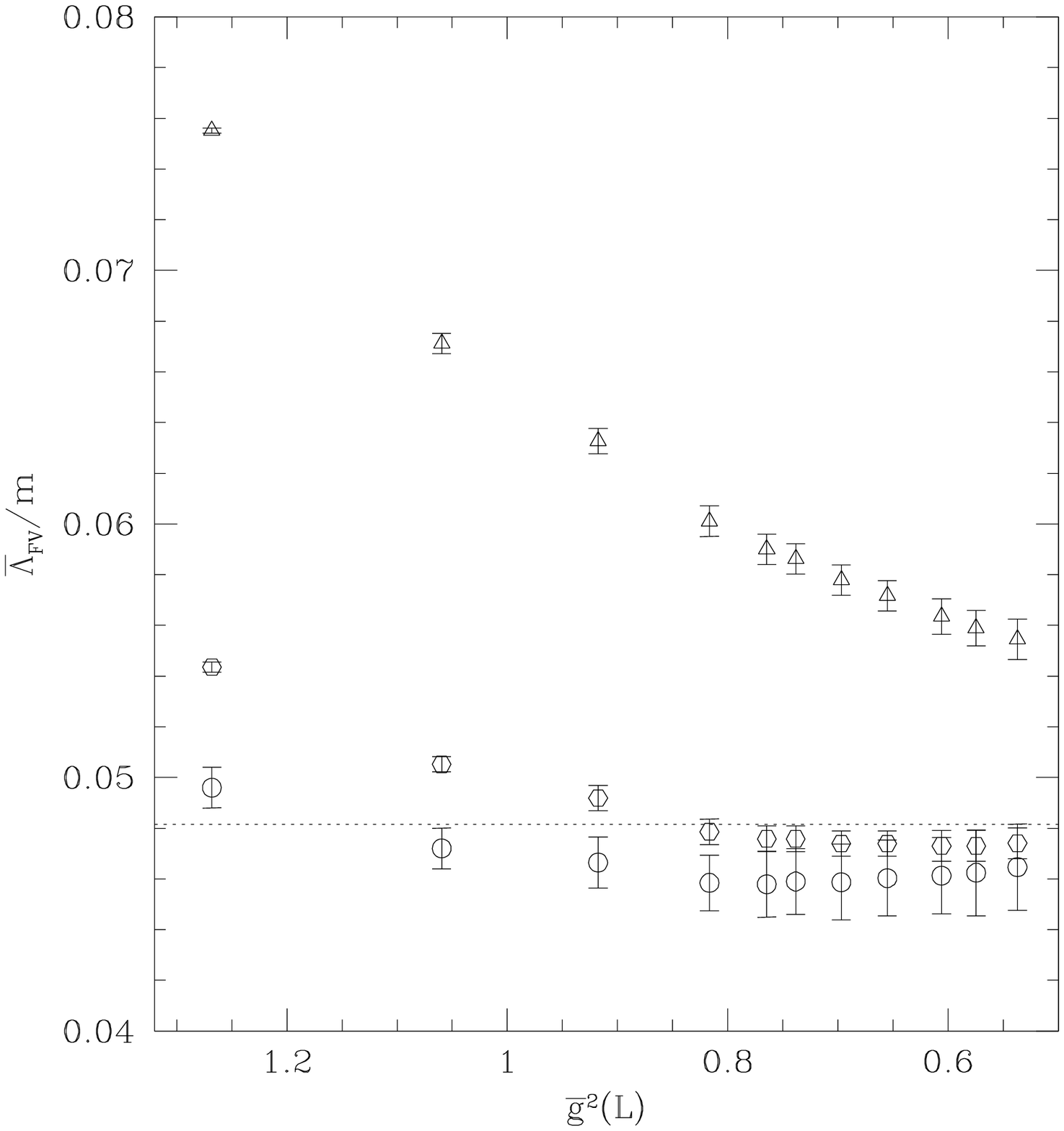}
\end{center}
\caption{
Two-, three- and four-loop approximations to $\Lambda_{FV}/m$ in the O(3) model
from Table~\ref{lambapm}.
The arrangement of the points and the dotted line is the same as 
in Fig.~\ref{fvml}.
}
\label{fvmlmo}
\end{figure}

Insertion of the data in Table~1 of Ref.~\cite{Sh2} in these equations
gives the numbers in Table~\ref{lambapm} which
are also plotted in Fig.~\ref{fvmlmo}.
In this case, we have exactly the same phenomenon as the former 
investigation.
Compared to the old figure with the four-loop $\beta$-function value
by Caracciolo et al.,
the four-loop curve for $\Lambda_{FV}/m$ at large couplings shifts
underneath in this case as well, 
so that the splitting of 3- and 4-loop curves becomes larger.
However, for small couplings which are actually our interesting region,
the difference between two figures,
one figure with the new $\beta$-function value by us
and the other with the old value by Caracciolo et al., vanishes.
Therefore, also in this investigation with the modified expansion to the
$\Lambda$-parameter,
we can conclude in the same way as the earlier investigation 
with the old value
of the $\beta$-function: all three curves converge to the value
predicted by Hasenfratz et al.

\begin{figure}[p]
\begin{center}
\leavevmode
\epsfxsize=122mm
\epsfbox{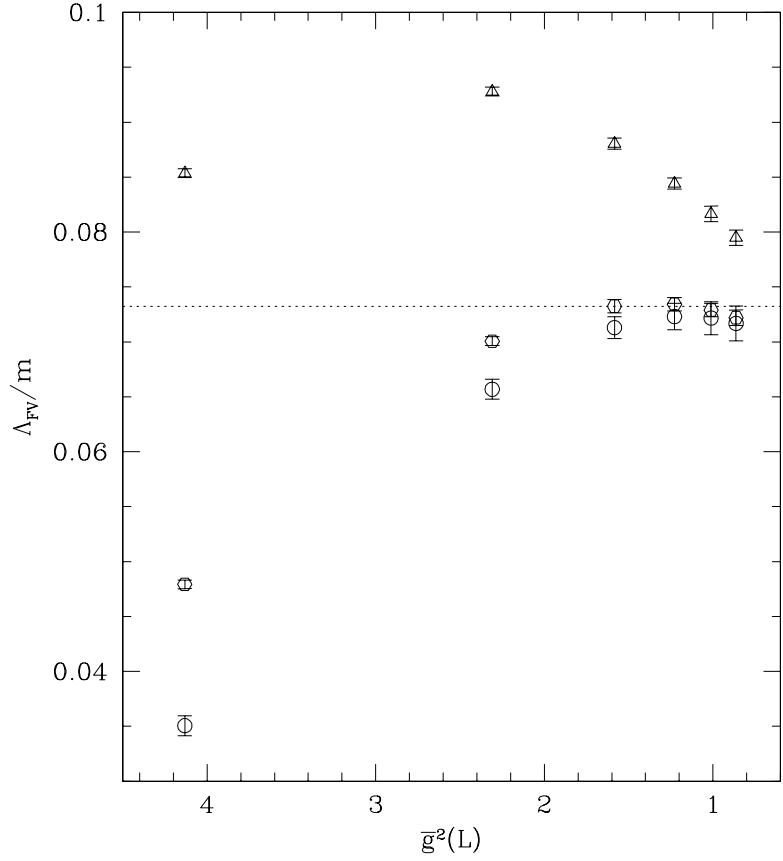}
\end{center}
\caption{
Two-, three- and four-loop approximations to $\Lambda_{FV}/m$ in the O(4) model
from Table~\ref{lambappo4f}.
The arrangement of the points is the same as in Fig.~\ref{fvml}.
The dotted line is the value of Hasenfratz et al. 
($\Lambda_{FV}/m=0.07321\cdots$).
}
\label{fvmlo4}
\end{figure}

We also want to discuss the changes of the diagrams
through the correction of the $\beta$-function in O(4) model.
By substituting the new value of the $\beta$-function in 
Eqs.~(\ref{2looplambdafv})-(\ref{4looplambdafv}),
we obtain the corrected results for $\Lambda_{FV}/m$ which are
written in Table~\ref{lambappo4f}
and also illustrated in Fig.~\ref{fvmlo4}.
As far as the modification of the four-loop curve is concerned,
Fig.~\ref{fvmlo4} shows exactly the same behavior as Fig.~\ref{fvml}:
as a result of the shift of the four-loop curve underneath, the order of 
3- and 4-loop curves changes, but for small couplings the difference 
between the old diagram and the new one vanishes
and we have the same argument with respect to the final conclusion.

\begin{figure}[p]
\begin{center}
\leavevmode
\epsfxsize=122mm
\epsfbox{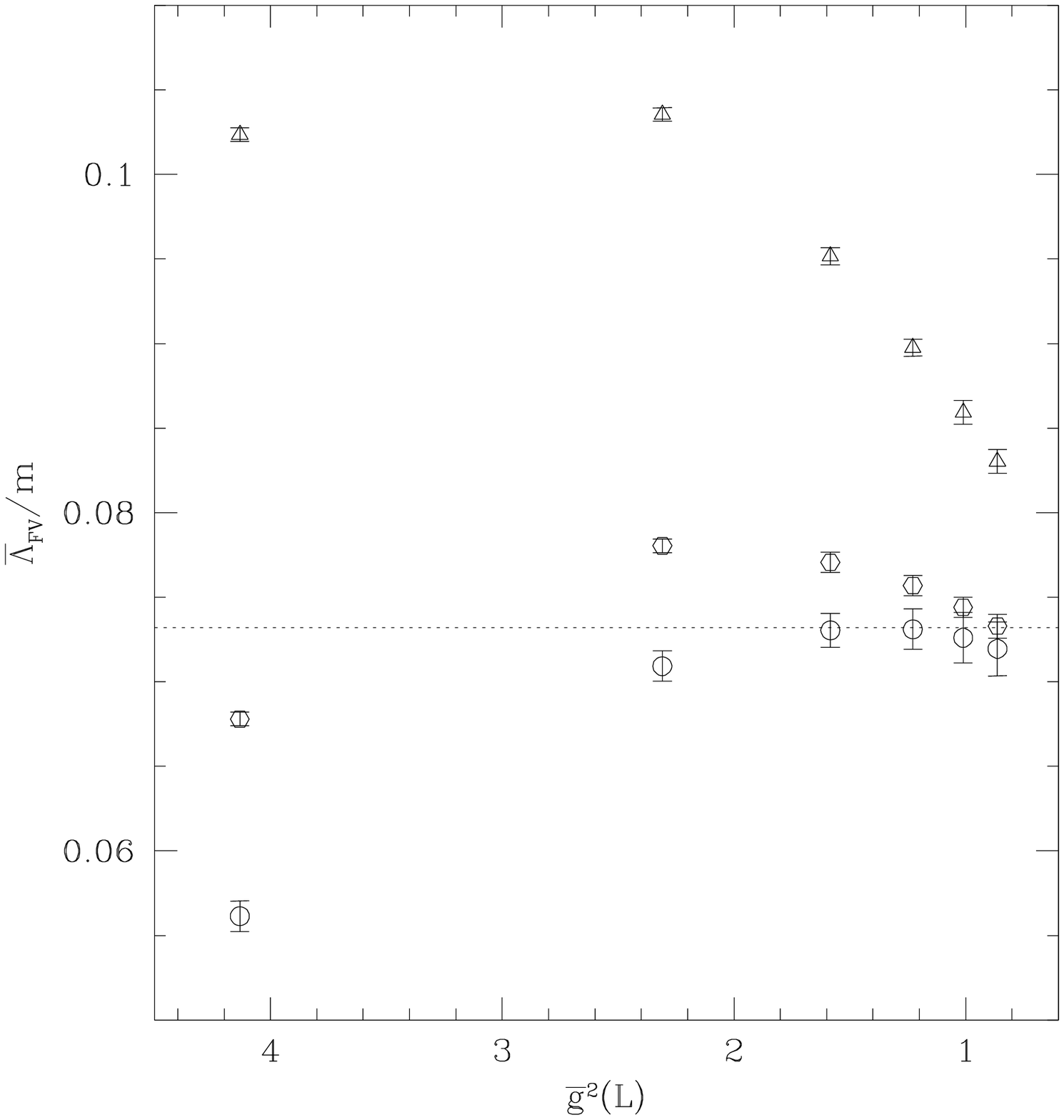}
\end{center}
\caption{
Two-, three- and four-loop approximations to $\Lambda_{FV}/m$ in the O(4) model
from Table~\ref{lambappo4s}.
The arrangement of the points and the dotted line is the same as 
in Fig.~\ref{fvmlo4}.
}
\label{fvmlo4m}
\end{figure}

As for the determination of 
$\Lambda_{FV}/m$ by means of the modified definition
of the perturbative $\Lambda_{FV}$ in O(4) model,
we obtain the results which are written in Table~\ref{lambappo4s} 
and also plotted in Fig.~\ref{fvmlo4m}.
In this case, we see exactly the same phenomenon as Fig.~\ref{fvmlmo}
which leads to the conclusion that for small couplings 
all three curves converge to the value by Hasenfratz et al. very well.

\begin{table}[p]\centering
\begin{tabular}{cccc}\hline
\hspace{0.5cm} \mbox{} $\bar{g}^2(L)$ \mbox{} \hspace{0.5cm} &
\hspace{0.5cm} \mbox{} $\bar{\Lambda}^{(2)}_{FV}/m$ \mbox{} \hspace{0.5cm} &
\hspace{0.5cm} \mbox{} $\bar{\Lambda}^{(3)}_{FV}/m$ \mbox{} \hspace{0.5cm} &
\hspace{0.5cm} \mbox{} $\bar{\Lambda}^{(4)}_{FV}/m$ \mbox{} \hspace{0.5cm} 
\\ \hline
 0.5372  & 0.0555(8) &  0.0474(6)  &  0.0465(17)  \\
 0.5747  & 0.0559(7) &  0.0473(6)  &  0.0462(17)  \\
 0.6060  & 0.0563(7) &  0.0473(6)  &  0.0461(15)  \\
 0.6553  & 0.0572(6) &  0.0474(5)  &  0.0460(15)  \\
 0.6970  & 0.0578(6) &  0.0474(5)  &  0.0459(15)  \\
 0.7383  & 0.0586(6) &  0.0476(5)  &  0.0459(23)  \\
 0.7646  & 0.0590(6) &  0.0476(5)  &  0.0458(13)  \\
 0.8166  & 0.0601(6) &  0.0479(5)  &  0.0458(11)  \\
 0.9176  & 0.0633(5) &  0.0492(5)  &  0.0466(10)  \\
 1.0595  & 0.0671(4) &  0.0505(3)  &  0.0472(8)\phantom{0}  \\
 1.2680  & 0.0755(1) &  0.0544(2)  &  0.0496(8)\phantom{0}
\\ \hline
\end{tabular}
\caption{\label{lambapm}
Alternative 2-, 3- and 4-loop approximations to $\Lambda_{FV}/m$
in the O(3) model}
\end{table}

\renewcommand{\arraystretch}{1.2}
\begin{table}[p]\centering
\begin{tabular}{cccc}\hline
\hspace{0.5cm} \mbox{} $\bar{g}^2(L)$ \mbox{} \hspace{0.5cm} &
\hspace{0.5cm} \mbox{} $\Lambda^{(2)}_{FV}/m$ \mbox{} \hspace{0.5cm} &
\hspace{0.5cm} \mbox{} $\Lambda^{(3)}_{FV}/m$ \mbox{} \hspace{0.5cm} &
\hspace{0.5cm} \mbox{} $\Lambda^{(4)}_{FV}/m$ \mbox{} \hspace{0.5cm} 
\\ \hline
 0.863  & 0.0795(7)  &    0.0722(7) &   0.0717(16) \\ 
 1.011  & 0.0817(7)  &    0.0729(6) &   0.0722(15) \\ 
 1.228  & 0.0844(5)  &    0.0734(6) &   0.0723(12) \\
 1.584  & 0.0880(5)  &    0.0733(6) &   0.0713(10) \\
 2.309  & 0.0928(4)  &    0.0701(4) &   0.0657(9)\phantom{0} \\
 4.132  & 0.0853(4)  &    0.0479(4) &   0.0351(9)\phantom{0} 
\\ \hline
\end{tabular}
\caption{\label{lambappo4f}
Two-, three- and four-loop approximations to $\Lambda_{FV}/m$
in the O(4) model}
\end{table}

\renewcommand{\arraystretch}{1.2}
\begin{table}[p]\centering
\begin{tabular}{cccc}\hline
\hspace{0.5cm} \mbox{} $\bar{g}^2(L)$ \mbox{} \hspace{0.5cm} &
\hspace{0.5cm} \mbox{} $\bar{\Lambda}^{(2)}_{FV}/m$ \mbox{} \hspace{0.5cm} &
\hspace{0.5cm} \mbox{} $\bar{\Lambda}^{(3)}_{FV}/m$ \mbox{} \hspace{0.5cm} &
\hspace{0.5cm} \mbox{} $\bar{\Lambda}^{(4)}_{FV}/m$ \mbox{} \hspace{0.5cm} 
\\ \hline
 0.863  & 0.0830(7) &   0.0733(7) &   0.0720(16) \\
 1.011  & 0.0859(7) &   0.0744(6) &   0.0726(15) \\
 1.228  & 0.0898(5) &   0.0757(6) &   0.0731(12) \\
 1.584  & 0.0952(5) &   0.0771(6) &   0.0730(10) \\
 2.309  & 0.1035(4) &   0.0780(4) &   0.0709(9)\phantom{0}\\
 4.132  & 0.1024(4) &   0.0678(4) &   0.0561(9)\phantom{0} 
\\ \hline
\end{tabular}
\caption{\label{lambappo4s}
Alternative 2-, 3- and 4-loop approximations to $\Lambda_{FV}/m$
in the O(4) model}
\end{table}

Summarizing the whole discussions up to now, we conclude that 
the correction of the four-loop $\beta$-function has an influence on our 
earlier investigation for the determination of 
the mass gap only at large couplings,
while in the actually interesting region of small couplings 
it provides practically no changes.
Therefore, our conclusions with the old value of the $\beta$-function
by Caracciolo et al. remain valid and we can say that 
in all of the investigated cases our determinations of the mass gap 
with the corrected value of the four-loop $\beta$-function by us
agree with those by Hasenfratz et al. very well.

\section*{Acknowledgement}

I greatly acknowledge Peter Weisz for many useful discussions
as well as for reading this manuscript.

\newpage
\begin{appendix}

\section{Evaluation of lattice Feynman diagrams}\label{appa}
\setcounter{equation}{0}

In this appendix we define the lattice integrals 
appearing in Table~\ref{4loopc}
of section~\ref{sec2} and show how they can be evaluated.

\subsection{Evaluation of $W_2$}

$W_2$ is defined as a finite part of an infrared divergent integral:
\eqb\label{w2}
W_2 = \lim_{h\to 0}\bigg[\overline{W}_2(h) 
-\frac16K(h)^3+\bigg(\frac18-\frac{1}{8\pi}\bigg)K(h)^2
-\bigg(\frac{1}{32}+\frac{1}{16\pi^2}-\frac{1}{16\pi}
-\frac{R}{2}\bigg)K(h)\bigg]
\eqe
with 
\eqb\label{w2b}
\overline{W}_2(h) = \int\limits_{qrst}
\frac{\sum_{\mu,\nu}
\sin q_{\mu}\sin r_{\nu}\sin s_{\mu}\sin t_{\nu}}
{(\hat q^2+h)(\hat r^2+h)(\hat s^2+h)(\hat t^2+h)}
\frac{1}{\widehat{(q+r)}^2+h}\bar\delta(q+r+s+t),
\eqe
where $\bar\delta(q)$ stands for a modified form of 
$\delta$-function $(2\pi)^2\delta(q)$ and
$\int_{q}$ for an abbreviated symbol of the integral sign
$\int_{-\pi}^{\pi}\frac{\md^2q}{(2\pi)^2}$.
The quantity $R$ is a finite integral 
which can be evaluated exactly
\eqb
R = \lim_{h\to 0}\bigg[h\int\limits_{qrs}
\frac{1}{(\hat q^2+h)(\hat r^2+h)(\hat s^2+h)}\bar\delta(q+r+s)\bigg]
=\frac{1}{24\pi^2}\psi'\bigg(\frac13\bigg)-\frac{1}{36}
\eqe
with $\psi(z)=\md\ln\Gamma(z)/\md z$,
while the function $K(h)$ is a divergent integral for \mbox{small $h$}
\eqb
K(h)=\int\limits_p\frac{1}{\hat p^2 + h} =
-\frac{1}{4\pi}\ln\frac{h}{32} + {\cal O}(h\ln h)
\eqe
with $\hat p^2\equiv\sum_{\mu}\hat p_{\mu}^2$, 
$\hat p_{\mu}\equiv 2\sin(p_{\mu}/2)$ and $\mu=0,1$.
Our aim is to evaluate the integral $\overline{W}_2(h)$ in the limit 
$h\to 0$ to confirm the logarithmic terms in Eq.~(\ref{w2}) and 
to determine the constant $W_2$ at the end.

Observing that 
Eq.~(\ref{w2b}) can be represented by the following symmetric form
\eqb\label{w2b1}
\overline{W}_2(h) = \int\limits_{p}
\frac{1}{\hat p^2+h}\sum_{\mu,\nu}F_{\mu\nu}(p,h)^2
\eqe
with
\eqb\label{gmnph}
F_{\mu\nu}(p,h) = \int\limits_{qr}
\frac{\sin q_{\mu}\sin r_{\nu}}{(\hat q^2+h)(\hat r^2+h)}\bar\delta(p+q+r),
\eqe
we try to 
evaluate Eq.~(\ref{w2b1}) by dividing into the following two integrals:
\eqb
\overline{W}_2(h) = I_1(h) + I_2(h),
\eqe
where
\eqnb\label{int1}
I_1(h) &=& \int\limits_{p}\frac{1}{\hat p^2+h}\sum_{\mu,\nu}F_{\mu\nu}(p,0)^2,
\\\label{int2}
I_2(h) &=& \int\limits_{p}\frac{1}{\hat p^2+h}
\sum_{\mu,\nu}\Big[F_{\mu\nu}(p,h)^2-F_{\mu\nu}(p,0)^2\Big].
\eqne

Considering, at first, the integral $I_1(h)$,
we transform the function $F_{\mu\nu}(p,0)$ to position space
to study how it behaves for small $p$:
\eqb\label{gmnp0}
F_{\mu\nu}(p,0) = -\frac14\sum_x\me^{ipx}x_{\mu}x_{\nu}H(x)^2,
\eqe
where $H(x)$ is defined by
\eqb\label{hx}
H(x) = \int\limits_{p}\me^{ipx}\ln\hat p^2 .
\eqe
For large $x$, Eq.~(\ref{hx}) shows the behavior 
$\lim\limits_{x\to\infty}H(x)\sim \frac{1}{x^2}$,
from which we can find out that 
$F_{\mu\nu}(p,0)$ diverges logarithmically for $p\to 0$.
We then break up Eq.~(\ref{gmnp0}) into converging and diverging part
\eqb\label{gmnexp}
F_{\mu\nu}(p,0) = -\frac14
\bigg[\bar F_{\mu\nu}(p,0)
+\frac{1}{4\pi}\sum_x\me^{ipx}
(\partial_{\mu}+\partial_{\mu}^*)(\partial_{\nu}+\partial_{\nu}^*)G(x)
-\frac{1}{2\pi}\delta_{\mu\nu}\sum_x\me^{ipx}H(x)\bigg],
\eqe
where 
the free propagator $G(x)$ including
its properties is discussed in detail in Ref.~\cite{Sh1} and
$\bar F_{\mu\nu}(p,0)$ is a well-defined function in the limit $p\to 0$,
constructed 
by subtracting from Eq.~(\ref{gmnp0}) the leading square terms in large $x$:
\eqb\label{gmnp0t}
\bar F_{\mu\nu}(p,0) = 
\sum_x \me^{ipx}\bigg[x_{\mu}x_{\nu}H(x)^2
-\frac{1}{4\pi}
(\partial_{\mu}+\partial_{\mu}^*)(\partial_{\nu}+\partial_{\nu}^*)G(x)
+\frac{1}{2\pi}\delta_{\mu\nu}H(x)\bigg].
\eqe
Next, we separate the $p=0$ part of Eq.~(\ref{gmnp0t})
to introduce a function $\tilde F_{\mu\nu}(p,0)$ which vanishes for $p=0$:
\eqb\label{gmnbp0}
\bar F_{\mu\nu}(p,0) = \tilde F_{\mu\nu}(p,0)+\delta_{\mu\nu}C
\eqe
with
\eqnb
\tilde F_{\mu\nu}(p,0) &=& 
\sum_x(\me^{ipx}-1)\bigg[x_{\mu}x_{\nu}H(x)^2
-\frac{1}{\pi}\partial_{\mu}\partial_{\nu}^*G(x)
+\frac{\delta_{\mu\nu}}{2\pi}H(x)\bigg], \\
C &=& 
\frac12\sum_x\bigg[x^2H(x)^2
-\frac{1}{4\pi}\sum_{\mu}(\partial_{\mu}+\partial_{\mu}^*)^2G(x)
+\frac{1}{\pi}H(x)\bigg].
\eqne
The numerical value of $C$, which can
be computed very precisely by applying the position space method~\cite{Sh1},
is $0.369899\cdots$.

Substituting Eq.~(\ref{gmnexp}), together with Eq.~(\ref{gmnbp0}), 
in Eq.~(\ref{int1}) and going back to momentum space, we have
\eqb\label{i1h}
I_1(h) = \frac{1}{16}\int\limits_{p}\frac{1}{\hat p^2+h}
\sum_{\mu,\nu}\bigg[\tilde F_{\mu\nu}(p,0)
-\frac{1}{\pi}\frac{\sin p_{\mu}\sin p_{\nu}}{\hat p^2}
-\frac{1}{2\pi}\delta_{\mu\nu}\ln\hat p^2
+\delta_{\mu\nu}C\bigg]^2
\eqe
with 
\eqb
\tilde F_{\mu\nu}(p,0) = -4\int\limits_{qr}
\frac{\sin q_{\mu}\sin r_{\nu}}{\hat q^2\hat r^2}\bar\delta(p+q+r)
+\frac{1}{\pi}\frac{\sin p_{\mu}\sin p_{\nu}}{\hat p^2}
+\frac{1}{2\pi}\delta_{\mu\nu}\ln\hat p^2-\delta_{\mu\nu}C .
\eqe
Expanding the square, Eq.~(\ref{i1h}) produces the following 
ten integrals which we can evaluate separately:
\eqnb
I_1(h) &=& \frac{1}{16}\bigg[2C^2K(h)-\frac{2}{\pi}CI_1^a(h)
+\frac{1}{2\pi^2}I_1^b(h)-\frac{2}{\pi}CI_1^c(h)+\frac{1}{\pi^2}I_1^d(h)
\nonumber\\ & &
+\frac{1}{\pi^2}I_1^e(h)+2CI_1^f(h)-\frac{1}{\pi}I_1^g(h)
-\frac{2}{\pi}I_1^h(h)+I_1^i(h)\bigg]
\eqne
with
\eqnb
I_1^a(h) &=& \int\limits_{p}\frac{\ln\hat p^2}{\hat p^2+h}, \\
I_1^b(h) &=& \int\limits_{p}\frac{(\ln\hat p^2)^2}{\hat p^2+h}, \\
I_1^c(h) &=& \int\limits_{p}
\frac{1}{\hat p^2+h}\frac{\sum_{\mu}\sin^2p_{\mu}}{\hat p^2}, \\
I_1^d(h) &=& \int\limits_{p}
\frac{\ln\hat p^2}{\hat p^2+h}\frac{\sum_{\mu}\sin^2p_{\mu}}{\hat p^2}, \\
I_1^e(h) &=& \int\limits_{p}
\frac{1}{\hat p^2+h}\sum_{\mu,\nu}
\frac{\sin^2p_{\mu}\sin^2p_{\nu}}{(\hat p^2)^2}, \\\label{i1fh}
I_1^f(h) &=& \int\limits_{p}
\frac{1}{\hat p^2+h}\sum_{\mu}\tilde F_{\mu\mu}(p,0), \\
I_1^g(h) &=& \int\limits_{p}
\frac{\ln\hat p^2}{\hat p^2+h}\sum_{\mu}\tilde F_{\mu\mu}(p,0), \\\label{i1hh}
I_1^h(h) &=& \int\limits_{p}
\frac{1}{\hat p^2+h}\sum_{\mu,\nu}\tilde F_{\mu\nu}(p,0)
\frac{\sin p_{\mu}\sin p_{\nu}}{\hat p^2}, \\\label{i1ih}
I_1^i(h) &=& \int\limits_{p}
\frac{1}{\hat p^2+h}\sum_{\mu,\nu}\tilde F_{\mu\nu}(p,0)^2.
\eqne
The first five integrals diverge for small $h$.
It is not difficult to first compute the divergent terms analytically
and then evaluate the remaining constant terms numerically 
in the limit $h\to 0$.
Our results are
\eqnb
I_1^a(h) &=& -\frac{1}{8\pi}(\ln h)^2+c_1+{\cal O}(h\ln h),\\
I_1^b(h) &=& -\frac{1}{12\pi}\Big[(\ln h)^3+\pi^2\ln h\Big]+c_2
+{\cal O}(h\ln h),\\
I_1^c(h) &=& K(h)-\frac{1}{\pi^2}c_3+{\cal O}(h\ln h),\\
I_1^d(h) &=& -\frac{1}{8\pi}(\ln h)^2+c_4+{\cal O}(h\ln h),\\
I_1^e(h) &=& K(h)+c_5+{\cal O}(h\ln h),
\eqne
where the values for $c_1,\cdots,c_5$ are listed in Table~\ref{w2cons}.
The last four integrals in Eqs.~(\ref{i1fh})-(\ref{i1ih}) 
are, on the other hand, well-defined for $h\to 0$.
Eqs.~(\ref{i1fh})-(\ref{i1hh}) can be transformed
to lattice sums in this limit
\eqnb\label{i1f0}
I_1^f(0) &=& \sum_xG(x)\bigg[x^2H(x)^2
-\frac{1}{4\pi}\sum_{\mu}(\partial_{\mu}+\partial_{\mu}^*)^2G(x)
+\frac{1}{\pi}H(x)\bigg],\\
I_1^g(0) &=& \sum_xL(x)\bigg[x^2H(x)^2
-\frac{1}{4\pi}\sum_{\mu}(\partial_{\mu}+\partial_{\mu}^*)^2G(x)
+\frac{1}{\pi}H(x)\bigg],\\\label{i1h0}
I_1^h(0) &=& -\frac14\sum_xG(x)\sum_{\mu,\nu}x_{\nu}
\Big[V_{\mu\nu}(x+\hat\mu)-V_{\mu\nu}(x-\hat\mu)\Big]+\frac{C}{4},
\eqne
where the functions $L(x)$ and $V_{\mu\nu}(x)$ are defined by
\eqnb
L(x) &=& \int\limits_p\frac{\ln\hat p^2}{\hat p^2}(\me^{ipx}-1),\\
V_{\mu\nu}(x) &=& x_{\mu}x_{\nu}H(x)^2
-\frac{1}{4\pi}
(\partial_{\mu}+\partial_{\mu}^*)(\partial_{\nu}+\partial_{\nu}^*)G(x)
+\frac{1}{2\pi}\delta_{\mu\nu}H(x).
\eqne
The numerical values for Eqs.~(\ref{i1f0})-(\ref{i1h0})
which we computed by applying the coordinate space technique
are given in Table~\ref{w2cons}.
As for the last integral $I_1^i(0)$ of Eq.~(\ref{i1ih}), 
we can compute it by a Fortran program calling NAGLIB integration subroutines.
We get a number in Table~\ref{w2cons}.

Let us now consider the integral $I_2(h)$ of Eq.~(\ref{int2}).
With help of the function
\eqb\label{hmnph}
H_{\mu\nu}(p,h) = F_{\mu\nu}(p,h)-F_{\mu\nu}(p,0),
\eqe
it can be divided into two integrals
\eqb
I_2(h) = A(h)+2B(h)
\eqe
with
\eqnb\label{ah1}
A(h) &=& \int\limits_{p}\frac{1}{\hat p^2+h}
\sum_{\mu,\nu}H_{\mu\nu}(p,h)^2, \\\label{bhs}
B(h) &=& \int\limits_{p}\frac{1}{\hat p^2+h}
\sum_{\mu,\nu}H_{\mu\nu}(p,h)F_{\mu\nu}(p,0)
\eqne
which we would like to evaluate in the $h\to 0$ limit.

Noting that Eq.~(\ref{ah1}) has a finite limit for $h\to 0$,
we perform the variable change $p\to\sqrt{h}q$ and take $h\to 0$
to transform to
\eqb\label{ahh0}
A(h)\stackrel{h\to 0}{=}
\int^{\infty}_{-\infty}\frac{\mbox{d}^2q}{(2\pi)^2}
\frac{1}{q^2+1}\sum_{\mu,\nu}\bar H_{\mu\nu}(q)^2
\eqe
with
\eqnb
\bar H_{\mu\nu}(q) &=& \lim_{h\to 0}H_{\mu\nu}(q\sqrt{h},h)\nonumber\\
\label{hmnq}
&=& \int^{\infty}_{-\infty}\frac{\mbox{d}^2q}{(2\pi)^2}r_{\mu}s_{\nu}
\bigg(\frac{1}{r^2+1}\frac{1}{s^2+1}-\frac{1}{r^2s^2}\bigg),
\hspace{0.2cm}s=q-r
\eqne
which can exactly be evaluated by applying Feynman parametrization
\eqb
\bar H_{\mu\nu}(q) = \frac{1}{8\pi}\delta_{\mu\nu}
\Bigg(-\ln q^2+\frac{2\sqrt{q^2+4}\,\,
{\rm arctanh}\sqrt{\frac{q^2}{q^2+4}}}{\sqrt{q^2}}
\Bigg)-\frac{q_{\mu}q_{\nu}}{\pi}
\frac{{\rm arctanh}\sqrt{\frac{q^2}{q^2+4}}}{(q^2)^{3/2}\sqrt{q^2+4}}.
\eqe
By transforming to polar coordinates,
the two-dimensional integral $A(0)$ of Eq.~(\ref{ahh0}) 
can be reduced to an one-dimensional
one being available to a numerical integration to high precision.
Its result is reported in Table~\ref{w2cons}.

The integral $B(h)$ from Eq.~(\ref{bhs})
is a little more tricky since it diverges for $h\to 0$.
We apply Eq.~(\ref{gmnexp}) to rewrite
\eqb
B(h) = -\frac14\Big[B_1(h) + B_2(h)\Big],
\eqe
where
\eqnb
B_1(h) &=& \int\limits_{p}\frac{1}{\hat p^2+h}
\sum_{\mu,\nu}H_{\mu\nu}(p,h)\tilde F_{\mu\nu}(p,0), \\
B_2(h) &=& \int\limits_{p}\frac{1}{\hat p^2+h}
\sum_{\mu,\nu}H_{\mu\nu}(p,h)
\bigg[-\frac{\delta_{\mu\nu}}{2\pi}\ln\hat p^2
-\frac{1}{\pi}\frac{\sin p_{\mu}\sin p_{\nu}}{\hat p^2}+\delta_{\mu\nu}C\bigg].
\eqne
For the evaluation of $B_1(h)$, we perform the variable change $p\to\sqrt{h}q$
and let $h\to 0$. Then, we  have
\eqb
B_1(h) \stackrel{h\to 0}{=} 
\int^{\infty}_{-\infty}\frac{\mbox{d}^2q}{(2\pi)^2}
\frac{1}{q^2+1}\sum_{\mu,\nu}\bar H_{\mu\nu}(q)\tilde F_{\mu\nu}(0,0)
\eqe
which is identically zero.
As for the integral $B_2(h)$, we do the same variable change $p\to\sqrt{h}q$.
In the limit $h\to 0$, we get
\eqb
B_2(h) \stackrel{h\to 0}{=} -\frac{1}{2\pi}(\ln h)B_2^a
-\frac{1}{2\pi}B_2^b-\frac{1}{\pi}B_2^c+CB_2^d
\eqe
with the two-dimensional, well-defined integrals
\eqnb
B_2^a &=& \int^{\infty}_{-\infty}\frac{\mbox{d}^2q}{(2\pi)^2}
\frac{1}{q^2+1}\sum_{\mu}\bar H_{\mu\mu}(q),\\
B_2^b &=& \int^{\infty}_{-\infty}\frac{\mbox{d}^2q}{(2\pi)^2}
\frac{\ln q^2}{q^2+1}\sum_{\mu}\bar H_{\mu\mu}(q),\\
B_2^c &=& \int^{\infty}_{-\infty}\frac{\mbox{d}^2q}{(2\pi)^2}
\frac{1}{q^2+1}\sum_{\mu,\nu}\bar H_{\mu\nu}(q)\frac{q_{\mu}q_{\nu}}{q^2},\\
B_2^d &=& \int^{\infty}_{-\infty}\frac{\mbox{d}^2q}{(2\pi)^2}
\frac{1}{q^2+1}\sum_{\mu}\bar H_{\mu\mu}(q).
\eqne
By transforming to polar coordinates,
all of these integrals can again be reduced to one-dimensional ones
whose numerical values are listed in Table~\ref{w2cons}.

\renewcommand{\arraystretch}{1.2}
\begin{table}[tbp]\centering
\begin{tabular}{c|c||c|c}\hline
Constants & Values & Constants & Values \\ \hline
 $c_1$ & 
  \hspace{0.2cm}\phantom{-}0.150666215804(1)\phantom{3382}\hspace{0.2cm} &   
 $I_1^h(0)$ &
  \hspace{0.2cm}-0.0165452709(1)\phantom{79587}\hspace{0.2cm} \\
 $c_2$  &  \phantom{-}0.420051822207(1)\phantom{3382} &   
 $I_1^i(0)$ & \phantom{-}0.0169938575(1)\phantom{79587}  \\
 $c_3$  &  \phantom{-}1.682002936874882(1)\phantom{3} &   
 $A(0)$ &     \phantom{-}0.001687475879587(1)\\
 $c_4$  &            -0.0355227164897(1)\phantom{382} &   
 $B_2^a$ &    \phantom{-}0.017838149659537(1) \\
 $c_5$  &            -0.2327464349513382(1) &   
 $B_2^b$ &              -0.006827009250335(1) \\
 $I_1^f(0)$ &        -0.066086583814(1)\phantom{3382} & 
 $B_2^c$ &    \phantom{-}0.001497591836898(1) \\
 $I_1^g(0)$ &        -0.07360068(1)\phantom{33823814} & 
 $B_2^d$ &    \phantom{-}0.017838149659537(1) \\
\hline
\end{tabular}
\caption{\label{w2cons}
Numerical values for constant terms appearing in the determination of $W_2$.}
\end{table}

\subsection{Evaluation of $W_1$}

$W_1$ is defined as a finite part of an infrared divergent integral:
\eqb\label{w1}
W_1 = \lim_{h\to 0}\bigg[\overline{W}_1(h) 
+\frac12K(h)^2-\frac{3}{8\pi}K(h)\bigg]
\eqe
with 
\eqb\label{w1b}
\overline{W}_1(h) = \int\limits_{qrs}
\frac{\sum_{\mu}\Big(\hat q_{\mu}^4-2\hat q_{\mu}^2\hat r_{\mu}^2\Big)}
{(\hat q^2+h)^2(\hat r^2+h)(\hat s^2+h)}\bar\delta(q+r+s) .
\eqe
Our aim is again to evaluate the integral 
$\overline{W}_1(h)$ in the limit $h\to 0$ 
to confirm the logarithmic terms in Eq.~(\ref{w1})
and to determine the constant $W_1$ at the end.

Eq.~(\ref{w1b}) can be transformed to a symmetric form
\eqb\label{w1bs}
\overline{W}_1(h) = \int\limits_{qrs}
\frac{\sum_{\mu}\hat q_{\mu}^2
\Big[\widehat{(r+s)}_{\mu}^2-\hat r_{\mu}^2-\hat s_{\mu}^2\Big]}
{(\hat q^2+h)^2(\hat r^2+h)(\hat s^2+h)}\bar\delta(q+r+s)
\eqe
which can, by expanding the numerator, be divided into two integrals
\eqb
\overline{W}_1(h) = 2T_1(h)-\frac12T_2(h)
\eqe
with
\eqnb\label{t1ho}
T_1(h) &=& \int\limits_{qrs}
\frac{\sum_{\mu}\hat q_{\mu}^2\sin r_{\mu}\sin s_{\mu}}
{(\hat q^2+h)^2(\hat r^2+h)(\hat s^2+h)}\bar\delta(q+r+s),\\\label{t2ho}
T_2(h) &=& \int\limits_{qrs}
\frac{\sum_{\mu}\hat q_{\mu}^2\hat r_{\mu}^2\hat s_{\mu}^2}
{(\hat q^2+h)^2(\hat r^2+h)(\hat s^2+h)}\bar\delta(q+r+s).
\eqne

Noting that Eq.~(\ref{t1ho}) can, with help of Eq.~(\ref{gmnph}),
be represented by 
\eqb\label{t1h}
T_1(h) = \sum_{\mu}\int\limits_{q}
\frac{\hat q_{\mu}^2}{(\hat q^2+h)^2}F_{\mu\mu}(q,h),
\eqe
we can evaluate it analogously to Eq.~(\ref{w2b1}).
We break up Eq.~(\ref{t1h}) into two parts
\eqb
T_1(h) = T_1^a(h)+T_1^b(h),
\eqe
where 
\eqnb\label{t1ah}
T_1^a(h) &=& \sum_{\mu}\int\limits_{q}
\frac{\hat q_{\mu}^2}{(\hat q^2+h)^2}F_{\mu\mu}(q,0), \\\label{t1bh}
T_1^b(h) &=& \sum_{\mu}\int\limits_{q}
\frac{\hat q_{\mu}^2}{(\hat q^2+h)^2}
\Big[F_{\mu\mu}(q,h)-F_{\mu\mu}(q,0)\Big].
\eqne

By analogy to $I_1(h)$ of Eq.~(\ref{int1}), Eq.~(\ref{t1ah}) can be expanded to
\eqb
T_1^a(h) = -\frac{1}{4}\sum_{\mu}\int\limits_{q}
\frac{\hat q_{\mu}^2}{(\hat q^2+h)^2}
\bigg[\tilde F_{\mu\mu}(q,0)
-\frac{1}{\pi}\frac{\sin^2 q_{\mu}}{\hat q^2}
-\frac{1}{2\pi}\delta_{\mu\nu}\ln\hat q^2+C\bigg]
\eqe
which produces the following four integrals being able to 
be evaluated separately:
\eqb
T_1^a(h) = -\frac{1}{4}\bigg[CJ_1(h)-\frac{1}{2\pi}J_2(h)-\frac{1}{\pi}J_3(h)
+J_4(h)\bigg]
\eqe
with
\eqnb
J_1(h) &=& \int\limits_q
\frac{\hat q^2}{(\hat q^2+h)^2},\\
J_2(h) &=& \int\limits_q
\frac{\hat q^2}{(\hat q^2+h)^2}\ln\hat q^2,\\
J_3(h) &=& \sum_{\mu}\int\limits_q
\frac{\hat q_{\mu}^2}{(\hat q^2+h)^2}\frac{\sin^2q_{\mu}}{\hat q^2},\\
J_4(h) &=& \sum_{\mu}\int\limits_q
\frac{\hat q_{\mu}^2}{(\hat q^2+h)^2}\tilde F_{\mu\mu}(q,0).
\eqne
The first three integrals diverge for $h\to 0$.
After having isolated the divergent parts analytically,
the remaining constant terms can be determined numerically:
\eqnb
J_1(h) &=& K(h)-\frac{1}{4\pi}+{\cal O}(h\ln h),\\
J_2(h) &=& -\frac{1}{8\pi}\Big[(\ln h)^2+2\ln h\Big]+k_1+{\cal O}(h\ln h),\\
J_3(h) &=& -\frac{3}{16\pi}\ln h+k_2+{\cal O}(h\ln h),
\eqne
where the values for $k_1$ and $k_2$ are given in Table~\ref{w1cons}.
The last integral $J_4(h)$ is, on the other hand, 
infrared finite and can, in the $h\to 0$ limit, be transformed to 
a lattice sum
\eqb
J_4(0) = -\sum_x\sum_{\mu}\Big[(\partial_{\mu}-\partial_{\mu}^*)G_2(x)\Big]
\bigg[x_{\mu}^2H(x)^2-\frac{1}{4\pi}(\partial_{\mu}+\partial_{\mu}^*)^2G(x)
+\frac{1}{2\pi}H(x)\bigg],
\eqe
where the function $G_2(x)$ including its properties is discussed in detail in
Ref.~\cite{Sh1}.
The number of $J_4(0)$ computed by the position space technique is shown
in Table~\ref{w1cons}.

Concerning the integral $T_1^b(h)$ of Eq.~(\ref{t1bh}),
it can be expressed by
\eqb\label{t2b}
T_1^b(h) = \sum_{\mu}\int\limits_q
\frac{\hat q_{\mu}^2}{(\hat q^2+h)^2}H_{\mu\mu}(q,h)
\eqe
with $H_{\mu\mu}(q,h)$ from Eq.~(\ref{hmnph}).
By changing the variable $q=\sqrt{h}p$ and taking the limit $h\to 0$,
Eq.~(\ref{t2b}) becomes
\eqb
T_1^b(0) = \sum_{\mu}
\int^{\infty}_{-\infty}\frac{\mbox{d}^2p}{(2\pi)^2}
\frac{p_{\mu}^2}{(p^2+1)^2}\bar H_{\mu\mu}(p).
\eqe
By transforming to polar coordinates, this two-dimensional
integral can be reduced to
one-dimensional one whose numerical value is reported in Table~\ref{w1cons}.

The last integral $T_2(h)$ of Eq.~(\ref{t2ho}) is to be evaluated by
dividing into two parts
\eqb
T_2(h) = T_2^a(h)+T_2^b(h),
\eqe
where 
\eqnb\label{t2ah}
T_2^a(h) &=& \int\limits_{qrs}\frac{\sum_{\mu}
\hat q_{\mu}^2\hat r_{\mu}^2\hat s_{\mu}^2}
{(\hat q^2+h)^2(\hat r^2+h)(\hat s^2+h)}
\Big[\bar\delta(q+r+s)-\bar\delta(r+s)\Big],\\\label{t2bh}
T_2^b(h) &=& \sum_{\mu}\int\limits_{q}
\frac{\hat q_{\mu}^2}{(\hat q^2+h)^2}
\bigg[\int\limits_{rs}
\frac{\hat r_{\mu}^2\hat s_{\mu}^2}{(\hat r^2+h)(\hat s^2+h)}
\bar\delta(r+s)\bigg].
\eqne
In the limit $h\to 0$, Eq.~(\ref{t2ah}) is finite 
and can be transformed to a lattice sum
\eqb
T_2^a(0) = -\sum_x\sum_{\mu}
\Big[(\partial_{\mu}-\partial_{\mu}^*)G_2(x)\Big]
\Big[(\partial_{\mu}-\partial_{\mu}^*)G(x)\Big]^2,
\eqe
while Eq.~(\ref{t2bh}) reduces to
\eqb
T_2^b(h) = \frac12k_3\bigg[K(h)-\frac{1}{4\pi}\bigg]+{\cal O}(h\ln h)
\eqe
with 
\eqb
k_3 = \int\limits_{r}\frac{1}{(\hat r^2)^2}\sum_{\mu}\hat r_{\mu}^4.
\eqe
The numerical values for $T_2^a(0)$ and $k_3$ are listed in Table~\ref{w1cons}.

\renewcommand{\arraystretch}{1.2}
\begin{table}[tbp]\centering
\begin{tabular}{c|c||c|c}\hline
Constants & Values & Constants & Values \\ \hline
 $k_1$  & \hspace{0.2cm} 0.150666215804(1)\phantom{30} \hspace{0.2cm} &   
 $J_4(0)$ & \hspace{0.2cm} -0.057000592735(1)\phantom{688} \hspace{0.2cm} \\
 $k_2$  & \hspace{0.2cm} 0.00658179485930(1) \hspace{0.2cm} &   
 $T_1^b(0)$ & \hspace{0.2cm} \phantom{-}0.001497591836898(1) \hspace{0.2cm} \\
 $k_3$  & \hspace{0.2cm} 0.68169011381621(1) \hspace{0.2cm} &   
 $T_2^a(0)$ & \hspace{0.2cm} -0.02533029591(1)\phantom{6898} \hspace{0.2cm} \\
\hline
\end{tabular}
\caption{\label{w1cons}
Numerical values for constant terms appearing in the determination of $W_1$.}
\end{table}

\subsection{Evaluation of the constant lattice integrals}

The constant lattice integrals appearing in Table~\ref{4loopc} 
of section~\ref{sec2} are defined as follows:
\eqsb
G_1 &=& -\frac14\int\limits_{qrs}
\frac{\hat{q}^2-\hat r^2-\hat s^2}{\hat r^2\hat s^2}
\frac{\sum_{\mu}\hat{q}_{\mu}^4}{(\hat{q}^2)^2}\bar\delta(q+r+s),\\
J &=& \int\limits_{qrst}
\frac{\Big(\sum_{\mu}\hat q_{\mu}\hat r_{\mu}\hat s_{\mu}\hat t_{\mu}
\Big)^2}{\hat q^2\hat r^2\hat s^2\hat t^2}\bar\delta(q+r+s+t),\\
K &=& \int\limits_{qrst}
\frac{\Big[\widehat{(q+r)}^2-\hat q^2-\hat r^2\Big]
\Big[\widehat{(s+t)}^2-\hat s^2-\hat t^2\Big]}
{\hat q^2\hat r^2\hat s^2\hat t^2}\bar\delta(q+r+s+t),\\
L_1 &=& \int\limits_{qrst}
\frac{\sum_{\mu,\nu}
\hat q_{\mu}\hat q_{\nu}\hat s_{\mu}\hat s_{\nu}\hat t_{\mu}\hat t_{\nu}
\Big[\hat r_{\mu}\hat r_{\nu}\widehat{(s+t)}^2-
\widehat{(s+t)_{\mu}}\widehat{(s+t)_{\nu}}\hat r^2\Big]}
{(\hat q^2)^2\hat r^2\hat s^2\hat t^2\widehat{(s+t)}^2}\bar\delta(q+r+s+t),\\
V_1 &=& \int\limits_{qrst}
\frac{\sum_{\mu}\hat q_{\mu}\hat r_{\mu}\hat s_{\mu}\hat t_{\mu}}
{\hat q^2\hat r^2\hat s^2\hat t^2}\bar\delta(q+r+s+t),\\
V_2 &=& \int\limits_{qrst}
\frac{\Big[\widehat{(s+t)}^2-\hat s^2-\hat t^2\Big]}{\hat s^2\hat t^2}
\sum_{\mu}
\frac{\Big[\frac12\widehat{(q+r)_{\mu}^2}\hat q^2
+\widehat{(q+r)^2}\hat q_{\mu}^2\Big]}{(\hat q^2)^2\hat r^2}\nonumber\\ & &
\times\frac{\Big[(\hat q_{\mu}^2+\hat r_{\mu}^2)\widehat{(q+r)^2}-
(\hat q^2+\hat r^2)\widehat{(q+r)_{\mu}^2}\Big]}
{[\widehat{(q+r)^2}]^2}\bar\delta(q+r+s+t),\\
V_3 &=& \int\limits_{qrst}
\frac{\Big[\widehat{(s+t)}^2-\hat s^2-\hat t^2\Big]
\Big[\widehat{(q+r)}^2-\hat q^2-\hat r^2\Big]}
{\hat q^2\hat r^2\hat s^2\hat t^2[\widehat{(q+r)^2}]^2}
\sum_{\mu}\widehat{(q+r)_{\mu}^4}\bar\delta(q+r+s+t),\\
V_4 &=& \int\limits_{qrst}
\frac{\sum_{\mu,\nu}\hat r_{\mu}^2\hat t_{\mu}^2\sin q_{\nu}\sin s_{\nu}}
{\hat q^2\hat r^2\hat s^2\hat t^2\widehat{(q+r)^2}}
\bar\delta(q+r+s+t),\\
V_5 &=& \int\limits_{qrst}
\frac{\sum_{\mu,\nu}\hat q_{\mu}\hat r_{\mu}\hat s_{\mu}\hat t_{\mu}
\sin^2 (q+r)_{\nu}}{\hat q^2\hat r^2\hat s^2\hat t^2\widehat{(q+r)^2}}
\bar\delta(q+r+s+t),\\
V_6 &=& \int\limits_{qrst}
\frac{\Big[\widehat{(s+t)}^2-\hat s^2-\hat t^2\Big]
\sum_{\mu}\widehat{(q+r)_{\mu}^2}\hat q_{\mu}\hat r_{\mu}
\cos\frac12(q+r)_{\mu}}
{\hat q^2\hat r^2\hat s^2\hat t^2\widehat{(q+r)^2}}
\bar\delta(q+r+s+t).
\eqse

We can transform all of these integrals except $V_4$ and $V_5$
to position space and then apply the coordinate space method~\cite{Sh1}:
\eqsb
G_1 &=& \frac18\sum_{x}\bigg\{\sum_{\mu}
\partial_{\mu}^*\partial_{\mu}\partial_{\mu}^*\partial_{\mu}G_2(x)\bigg\}
\bigg\{\sum_{\mu}\Big[
\Big((\partial_{\mu}-\partial_{\mu}^*)G(x)\Big)^2+
\Big((\partial_{\mu}+\partial_{\mu}^*)G(x)\Big)^2\Big]\bigg\},\\
J &=& \sum_{x}\sum_{\mu,\nu}
\Big[\partial_{\mu}\partial_{\nu}G(x)\Big]^4,\\
K &=& \frac14\sum_{x}\bigg\{\sum_{\mu}\Big[
\Big((\partial_{\mu}-\partial_{\mu}^*)G(x)\Big)^2+
\Big((\partial_{\mu}+\partial_{\mu}^*)G(x)\Big)^2\Big]\bigg\}^2,\\
L_1 &=& \sum_x\sum_{\mu,\nu}
\Big[\partial_{\mu}\partial_{\nu}G_2(x)\Big]
\Big[\partial_{\mu}\partial_{\nu}G(x)\Big]^3,\\
V_1 &=& \sum_{x}\sum_{\mu}\Big[\partial_{\mu}G(x)\Big]^4,\\
V_2 &=& -\frac12\sum_x\bigg\{
\sum_{\mu}\Big[(\partial_{\mu}-\partial_{\mu}^*)G(x)\Big]^2+
\sum_{\mu}\Big[(\partial_{\mu}+\partial_{\mu}^*)G(x)\Big]^2\bigg\}
\nonumber\\& &
\times\bigg\{G(x)\sum_{\mu}
(\partial_{\mu}-\partial_{\mu}^*)^2G_2(x) +
\sum_{\mu}\Big[(\partial_{\mu}-\partial_{\mu}^*)G(x)\Big]
\Big[(\partial_{\mu}-\partial_{\mu}^*)G_2(x)\Big]\bigg\},\\
V_3 &=& \sum_xG(x)^2
\sum_{\mu}(\partial_{\mu}-\partial_{\mu}^*)^2G(x)^2,\\
V_6 &=& \frac18\sum_xG(x)^2
\sum_{\mu}\Big\{
2\Big[(\partial_{\mu}-\partial_{\mu}^*)(\partial_{\mu}
+\partial_{\mu}^*)G(x)\Big]
\Big[G(x+2\hat\mu)-G(x-2\hat\mu)\Big]\nonumber\\& &
+\Big[(\partial_{\mu}+\partial_{\mu}^*)^2G(x)\Big]
\Big[(\partial_{\mu}-\partial_{\mu}^*)^2G(x)+
(\partial_{\mu}+\partial_{\mu}^*)^2G(x)\Big]\Big\}.
\eqse

Concerning the integral $V_4$, 
after transforming to a symmetric form
\eqb
V_4 = -\int\limits_t\frac{1}{\hat t^2}\sum_{\mu,\nu}A_{\mu\nu}(t)^2
\eqe
with
\eqb
A_{\mu\nu}(t) = \int\limits_q
\frac{\widehat{(t+q)_{\mu}^2}\sin q_{\nu}}{\widehat{(t+q)^2}\hat q^2},
\eqe
we can compute it by a Fortran program calling NAGLIB integration subroutines.
The integral $V_5$ can, on the other hand, be divided into
\eqb
V_5 = V_5^a-\frac14V_5^b,
\eqe
where
\eqnb
V_5^a &=& \int\limits_{qrst}
\frac{\sum_{\mu}\hat q_{\mu}\hat r_{\mu}\hat s_{\mu}\hat t_{\mu}}
{\hat q^2\hat r^2\hat s^2\hat t^2}\bar\delta(q+r+s+t),\\
V_5^b &=& \int\limits_{qrst}
\frac{\sum_{\mu}\hat q_{\mu}\hat r_{\mu}\hat s_{\mu}\hat t_{\mu}
\sum_{\nu}\widehat{(q+r)_{\nu}^4}}
{\hat q^2\hat r^2\hat s^2\hat t^2\widehat{(q+r)^2}}
\bar\delta(q+r+s+t).
\eqne
$V_5^a$ is then able to be represented in coordinate space
\eqb
V_5^a = \sum_{x}\sum_{\mu}\Big[\partial_{\mu}G(x)\Big]^4,
\eqe
while $V_5^b$ can be symmetrized as
\eqb
V_5^b = \int\limits_{t}
\frac{\sum_{\mu}\hat t_{\mu}^4}{\hat t^2}
\sum_{\nu}B_{\nu}(t)^2
\eqe
with
\eqb
B_{\nu}(t) = \int\limits_{q}
\frac{\widehat{(t+q)_{\nu}}\hat q_{\nu}}{\widehat{(t+q)^2}\hat q^2}
\eqe
which is again to be evaluated by using NAGLIB integration subroutines.

\end{appendix}

\end{document}